\documentclass[prd,aps,preprint,nofootinbib,superscriptaddress]{revtex4-1}

\usepackage{amsmath}
\usepackage{epsfig}

\begin{document}


\title{Closing in on Supersymmetric Electroweak Baryogenesis\\ with Dark Matter Searches and the Large Hadron Collider}

\author{Jonathan Kozaczuk}
\email{jkozaczu@ucsc.edu}\affiliation{Department of Physics, University of California, 1156 High St., Santa Cruz, CA 95064, USA}

\author{Stefano Profumo}
\email{profumo@scipp.ucsc.edu}\affiliation{Department of Physics, University of California, 1156 High St., Santa Cruz, CA 95064, USA}\affiliation{Santa Cruz Institute for Particle Physics, Santa Cruz, CA 95064, USA}

\date{\today}

\begin{abstract}
\noindent We study the impact of recent direct and indirect searches for particle dark matter on supersymmetric models with resonant neutralino- or chargino-driven electroweak baryogenesis (EWB) and heavy sfermions. We outline regions of successful EWB on the planes defined by gaugino and higgsino mass parameters, and calculate the portions of those planes excluded by dark matter search results, and the regions soon to be probed by current and future experiments. We conclude that dark matter searches robustly exclude a wino-like lightest supersymmetric particle in successful EWB regions. Bino-like dark matter is still a possibility, although one that will be probed with a modest improvement in the sensitivity of current direct and indirect detection experiments. We also calculate the total production cross section of chargino and neutralino pairs at the Large Hadron Collider, with a center of mass energy of 7 and 14 TeV.
\end{abstract}

\maketitle

\section{Introduction}
Given the Standard Model (SM) of elementary particle physics, the matter content we observe in the universe today is quite preposterous: the universe should simply be filled with relic photons and neutrinos.  The existence of a net baryon number, and thus of ``ordinary'' matter, and the existence and dominance of non-baryonic cold dark matter (CDM) in the matter density budget manifestly point to physics beyond the Standard Model, and pose severe model building challenges. With the advent of the Large Hadron Collider (LHC) and with a diverse and far-ranging campaign to search for the particle(s) composing the dark matter, rapid progress in our understanding of these issues might be just around the corner.

As we review in some detail in Sec.~\ref{sec:asymmetry}, electroweak baryogenesis (EWB) is a compelling scenario in which the baryon asymmetry of the universe (BAU) is produced at the electroweak phase transition (EWPT) via particle scattering off of the expanding walls of bubbles of broken EW phase (see e.g. \cite{Huet:1995sh}). In its supersymmetric incarnation within the minimal supersymmetric extension of the Standard Model (MSSM), EWB naturally posits a dark matter candidate -- the lightest neutralino -- across wide portions of its parameter space. From a phenomenological standpoint, since EWB relies on selected particle species being in thermal equilibrium at the time of the EWPT, the mass scale of these particles must necessarily lie not far from the EW scale itself. As a result, any generic successful EWB model, and specifically supersymmetric scenarios, features relatively light, sub-TeV particles that would be copiously produced at the LHC, with the lightest neutralino typically quite light and with a sizable higgsino fraction \cite{EWB_and_EDMs, EWB_and_DM}. The latter particle properties typically imply very large dark matter direct and indirect detection rates, although this depends on the specifics of the MSSM parameter space points.

The particle spectrum of successful and phenomenologically viable MSSM EWB models is relatively simple. First, sfermions, with the possible exception of the right-handed stop, must be extremely heavy, compared to the electroweak scale, in order to avoid excessively large electric dipole moments. Second, for the class of CP-violating sources we consider here, charginos and neutralinos must have electroweak-scale masses, and the higgsino mass term $\mu$ must be close to one of the electroweak gaugino masses $M_1$ or $M_2$ (resonant neutralino or chargino-driven EW baryogenesis). The collider and dark matter phenomenology of a given EWB model then depends strongly on the relative hierarchy between the hypercharge and the $SU(2)$ gaugino soft supersymmetry breaking masses. For the sake of illustration, we employ here two paradigmatic choices --- one motivated by gaugino mass unification at the grand unification scale with $M_1<M_2$ at the electroweak scale, and one motivated by anomaly mediated supersymmetry breaking with $M_2<M_1$.

Searches for particle dark matter have seen significant and rapid progress, warranting close scrutiny of their impact on models residing in the regions of parameter space being explored, such as supersymmetric resonant chargino- or neutralino-driven EWB.  The potential of DM searches to constrain the EWB parameter space has been discussed previously in Ref.~\cite{EWB_and_DM}, however given recent developments we are now in a position to make more definitive statements about which parts of these regions are still an option for EWB and which are experimentally disfavored or ruled out.  At the same time, as the first LHC results on searches for supersymmetry with 1 fb${}^{-1}$ of integrated luminosity are being presented, outlining prospects for the discovery of this class of models with the LHC seems also particularly timely and compelling.

We begin this study in Sec.~\ref{sec:asymmetry} by outlining, on the relevant parameter space, the regions compatible with the successful production of the observed baryon asymmetry via electroweak baryogenesis, taking into account some of the model uncertainties (such as the bubble wall velocity); we then explore the dark matter particle phenomenology of models with $M_1<M_2$ (Sec.~\ref{sec:bino}) and $M_2<M_1$ (Sec.~\ref{sec:wino}), including the lightest neutralino relic density and the impact of recent direct and indirect searches (the latter via neutrino and gamma-ray telescopes). Finally, in Sec.~\ref{sec:lhc} we study the production cross section for all possible neutralino/chargino pairs at the LHC, and comment on the prospects of discovery for this class of models.  Sec.~\ref{sec:disc} contains our discussion and concluding remarks.

\section{The Baryon Asymmetry}\label{sec:asymmetry}
In the EWB scenario, $B+L$-violating weak sphaleron processes produce a net baryon number from an initial non-zero left-handed fermion doublet density, $n_L$, generated by CP-violating interactions of the various (s)particles in the thermal bath with the electroweak phase transition bubble wall \cite{Huet:1995sh}.  These sphaleron processes are in equilibrium in the unbroken phase but Boltzmann-suppressed in the broken phase, causing the baryon asymmetry to be frozen in once it passes into the broken phase, given a sufficiently strongly first-order phase transition.  The left-handed density $n_L$ is created as a result of the scattering of particles off of the spacetime-dependent Higgs vacuum expectation values (vevs) in the wall, leading to quantum transport equations (QTEs) for the relevant species involved.  These processes have been studied in depth in several different contexts (see e.g. Refs.~\cite{Huet:1995sh, EWB_and_EDMs, EWB_and_DM, Resonant_Relaxation, Chung:2008aya, Lepton_Mediated, Supergauge, Including_Yukawa, Carena:2002ss, Carena:2008vj, Konstandin:2004gy, Konstandin:2003dx, More_Relaxed, Balazs:2004ae, Huber:2001xf, CPV_and_EWB, Cline:2000kb}).

\subsection{MSSM Parameter Space}\label{sec:param}
We are interested in regions of the MSSM parameter space that produce a baryon-to-entropy ratio, $Y_B=n_B/s$, consistent with the WMAP 7-year data \cite{Komatsu:2010fb}, $Y_B^{WMAP}=(8.1-9.4) \times 10^{-11}$.  To calculate $Y_B$ as a function of the MSSM parameters, we follow closely the methods detailed in Refs. \cite{Resonant_Relaxation, Chung:2008aya, Lepton_Mediated, Supergauge, Including_Yukawa} in which the authors compute self-consistently both the CP-violating sources active in the bubble wall, to leading non-trivial order in the background spacetime-varying Higgs field, and the CP-conserving particle number-changing interactions that serve to relax the left-handed particle asymmetry.  The authors of Ref. \cite{Resonant_Relaxation} considered the sources corresponding to higgsino-gaugino scattering and stop mass mixing in the wall, however the latter was found to be suppressed by a factor of $\sim 10^{-3}$ relative to the higgsino source in the region of the MSSM where $m^2_{\widetilde{t}_R}\ll m^2_{\widetilde{t}_L}$ as is the case for our analysis, as discussed below.  In what follows we thus neglect the contribution of the stop source to $Y_B$.  In addition to the higgsino and stop sources, it was pointed out in Refs. \cite{Chung:2008aya, Lepton_Mediated, Supergauge} that sbottom and stau CP-violating  sources and CP-conserving relaxation rates can become active for a large ratio of Higgs vevs, $\tan\beta$, and small $m_{\widetilde{b}}$, $m_{\widetilde{\tau}}$.  The sbottom/stau Yukawa interactions tend to suppress the overall baryon asymmetry, however we will work with the assumption that $m_{\widetilde{b}}$, $m_{\widetilde{\tau}}\gg m_{\widetilde{t}_R}$ so that we can neglect these contributions as well.  This choice consequently weakens the dependence of the transport dynamics, and hence the overall baryon asymmetry, on $\tan\beta$ \cite{Supergauge}. 

It should be noted that the CP-violating sources used in the current study to calculate the BAU are not suppressed for large values of $\tan \beta$.  Given our simplified bubble wall profile, discussed below, the CP-violating sources we consider \cite{Resonant_Relaxation} (which are the dominant sources near resonance) are dependent on $\tan \beta$ only through the parameter $\Delta \beta$, as discussed in Sec.~\ref{sec:wall_param}.  We do not attempt to precisely account for this dependence as doing so correctly would require solving for the exact EWPT bubble profile which is beyond the scope of the present study.  Some other studies have suggested a suppression of the CP-violating sources, and hence the BAU, for large $\tan \beta$ (see, e.g.~\cite{Carena:2002ss, Carena:2008vj} and related work), especially off-resonance, however we do not expect such behavior to significantly affect our results.  We find that our calculations of the BAU are largely insensitive to $\tan \beta$ for values $\gtrsim 10$, as are the DM detection rates, thus we expect our conclusions to hold even if restricted to smaller values of $\tan \beta$ than that considered here.    For concreteness, we typically choose $\tan \beta=40$ as representative of values producing a maximal $Y_B$.  As with the other parameters discussed below, we take the resulting BAU to represent a conservative bound on the MSSM parameter space viable for EWB.  Additionally, the methods used to calculate the sources are also likely to overestimate $Y_B$ \cite{Carena:2002ss, Konstandin:2004gy, Konstandin:2003dx}, which is also consistent with the interpretation of our curves as an optimistic scenario for EWB.

For simplicity, we choose also to work in the so-called superequilibrium regime, in which supergauge interactions that convert a non-zero superpartner density to a density of the corresponding SM fermions are fast compared to the other particle number-changing processes.  This assumption allows us to write the QTEs in terms of the supermultiplet densities $Q$, $T$,  $H$ corresponding to the left-handed third generation (s)quarks, right-handed (s)top, and Higgs/higgsino densities, respectively.  The conditions for supergauge equilibrium are generally satisfied in the MSSM region of interest \cite{Supergauge}, although for other non-minmal SUSY models one may have to solve the full set of Boltzmann equations to accurately compute the baryon asymmetry.  We leave such an investigation to future work.

Given the assumptions above, the relevant CP-violating phases in calculating $Y_B$ are those arising in the higgsino-gaugino sector from the corresponding soft-breaking terms in the Lagrangian.  We assume universal CP-violating phases, so that \begin{equation} \phi_{\mu}=\arg(M_1\mu b^*)=\arg(M_2 \mu b^*) \end{equation} where $M_1$, $M_2$ are the bino, wino soft SUSY-breaking masses, $\mu$ is the higgsino-Higgs mass parameter, and $b$ is a soft SUSY-breaking Higgs mass parameter \footnote{Here, and throughout this manuscript, masses are understood to be given at the electroweak scale.}.  Relaxing this assumption to the case of non-universal CP-violating phases may be necessary to produce the observed baryon asymmetry if the next generation of EDM experiments yield null results, especially for large values of $\tan \beta$ \cite{EWB_and_EDMs}.

\begin{figure*}[!t]
\mbox{\includegraphics[width=0.55\textwidth,clip]{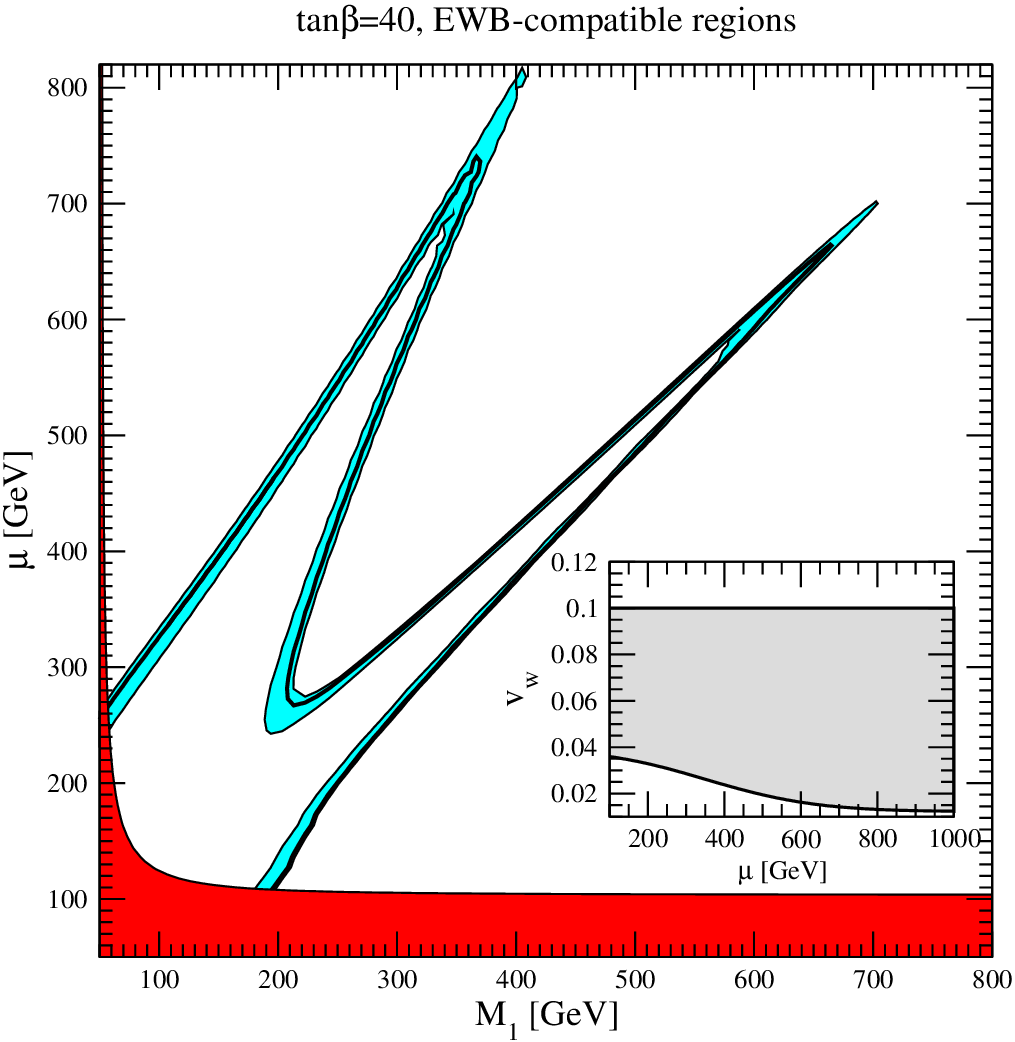}}
\caption{\label{fig:tb40_ewb}\it\small Regions compatible with resonant chargino-neutralino electroweak baryogenesis, on the $(M_1,\mu)$ plane at $\tan\beta=40$, for maximal gaugino-higgsino CP-violating phase $\sin\phi_\mu=1$ and for $m_A=300$ GeV. The cyan region corresponds to the band in the wall velocity $v_w$ shown in the inset.  The red shaded region is excluded by LEP searches for light neutralinos/charginos \cite{Amsler:2008zzb}}
\end{figure*}

Throughout our analysis we take all sfermions, other than the gauginos, higgsinos, and right-handed stops, to be heavy, $m_{sf} \sim 1-10$ TeV.  The gaugino masses, which are generally assumed to be unified to a common mass at some higher scale, typically organize themselves into patterns given the mechanism of supersymmetry breaking.  We follow Ref.~\cite{EWB_and_DM} and concentrate for the sake of illustration on two scenarios for our investigation, namely gravity-mediated SUSY breaking, which yields the pattern $M_1\approx M_2/2$, and anomaly-mediated SUSY breaking models (AMSB), for which $M_2\approx M_1/3$  \cite{Baer:2000gf}.  These hierarchies and their implications are discussed in more detail in Sec.~\ref{sec:bino} and Sec.~\ref{sec:wino}.  In calculating the baryon asymmetry we typically vary the lightest gaugino mass and $\mu$, focusing on the $(M_1,\mu)$ plane for the gravity-mediated case and the $(M_2,\mu)$ plane for the anomaly-mediated scenario.     

We calculate the baryon asymmetry numerically for values of $\mu$, $M_{1,2}$ between $100$ GeV and $1$ TeV to find parametric regions producing the observed\footnote{The uncertainty in $Y_B^{WMAP}$ leads to negligible uncertainties in the higgsino-gaugino mass planes.  We thus adopt the value $Y_B^{WMAP}=9.1\times 10^{-11}$ for consistency with previous studies.} baryon-to-entropy ratio for a given value of the CP-violating phase $\phi_{\mu}$.  Our results in the $(M_1,\mu)$ plane for $\sin\phi_{\mu}=1$, $\tan\beta=40$ are shown in Fig.\ref{fig:tb40_ewb}.  As we are considering the resonant EWB scenario \cite{Resonant_Relaxation, More_Relaxed}, the baryon asymmetry is generated primarily by CP-violating trilinear scalar sources near resonance.  For the higgsino-gaugino-vev interactions this resonant behavior occurs for nearly degenerate gaugino, higgsino masses, $M_{1,2}\sim\mu$, leading to the two-funnel structure in Fig.\ref{fig:tb40_ewb} as discussed in Refs.~\cite{EWB_and_EDMs, EWB_and_DM}.  

Since the gauginos in this scenario are light, it was previously realized that much of the parameter space suitable for successful EWB is also that required for a neutralino lightest supersymmetric particle (LSP), thus providing a viable dark matter candidate along with the baryon asymmetry (see e.g. Refs.~\cite{EWB_and_DM, EWB_and_EDMs, Balazs:2004ae}).  As a result, one can place further constraints on the gaugino-higgsino parameter space from the results produced by various dark matter searches.  In the resonant EWB scenario, however, the right-handed stop is typically also light to satisfy the requirement of a strongly first-order EWPT, and consequently to prevent the washout of the baryon asymmetry in the broken phase \cite{Laine:1998qk, Laine:2000xu, Carena:2008rt}.  A right-handed stop mass $m_{\widetilde{t}_R}$ of $\mathcal{O}(100$ GeV$)$ would imply significant regions of the gaugino-higgsino parameter space in which the RH stop is instead the LSP.  Following the strategy in Refs.~\cite{EWB_and_EDMs, EWB_and_DM} we expand the region of the parameter space simultaneously viable for EWB and a neutralino LSP by assuming that some other mechanism is responsible for ensuring a strongly first order phase transition.  Several such mechanisms have been proposed \cite{Heavy_Stops, Profumo:2007wc} that are decoupled from the mechanisms driving EWB and thus allowing for a heavy RH stop with  $m_{\widetilde{t}_R}\gtrsim 1$ TeV while still preventing baryon number washout\footnote{Such a mechanism may also be necessary in light of recent studies indicating that magnetic fields may be produced at the EWPT which might lower the sphaleron energy and thus necessitate a stronger phase transition to prevent baryon number washout \cite{DeSimone:2011ek}.}.  For example, extending the scalar sector of the theory by including (well-motivated) gauge singlets can augment the strength of the EWPT (see e.g. \cite{Profumo:2007wc} and references therein).  For increasing values of $m_{\widetilde{t}_R}$, the baryon asymmetry is suppressed  and our numerical results find no viable parameter space for EWB given $m_{\widetilde{t}_R} \gtrsim 500$ GeV.  This suppression is due to the erasure of chiral charges by strong sphaleron processes \cite{Giudice:1993bb} as discussed in the following subsection.   For heavier RH stop masses and large values of $\tan\beta$ one might circumvent this problem by taking relatively light sbottoms, staus ($\sim 1$ TeV) , a combination of which may serve to bolster the baryon asymmetry (albeit with a potential sign change) \cite{Chung:2008aya, Lepton_Mediated, Supergauge} and possibly extend the range of $m_{\widetilde{t}_R}$ consistent with the observed baryon-to-entropy ratio.  For the purposes of our numerical analysis, however, we take the RH SUSY breaking scalar mass $m_{\widetilde{U_3}}=0$, corresponding, for the value of the tri-scalar coupling, $\tan\beta$ and range of $\mu$ we consider, to a physical RH stop mass $m_{\widetilde{t}_R}\approx 160$ GeV, and consider our results to be a conservative outline of the gaugino-higgsino parameter space consistent with EWB and a neutralino LSP, deferring a detailed analysis of the more realistic case with finite sbottom and stau masses to future work.

The other MSSM parameters relevant to our analysis are chosen to satisfy both the condition of a strongly first-order phase transition and bounds from precision electroweak measurements \cite{Huet:1995sh, Huber:2001xf, Precision_EW1}.  We take the heavy Higgs mass to be set by the common scalar soft SUSY-breaking mass at the TeV scale, while for the light Higgs we take $m^2 _{H_u}=-(50)^2$ GeV$^2$.  To ensure that the lightest stop is mostly right handed, as required to prevent significant contributions to the $\rho$ parameter \cite{EWB_and_DM}, the mixing must be small and we take $A_t=200$ GeV\footnote{Note that precision electroweak and Higgs constraints also require $\tan\beta\gtrsim 5$ in the resonant EWB scenario \cite{Precision_EW1}}.  The baryon asymmetry depends sensitively on the mass of the CP-odd Higgs, $m_A$, through the quantity $\Delta \beta$ (which we discuss in the following section) and is suppressed for large values \cite{EWB_and_EDMs, EWB_and_DM}.  We generally take $m_A=300$ GeV and 1000 GeV as these two values bracket the interesting range for MSSM resonant EWB; namely, the low value corresponds to contributions to, for example, the branching ratio $b\to s\gamma$ close to the experimental limit, while the high value corresponds to a rather large suppression of the BAU.  This suppression may be alleviated by including additional non-resonant source terms in our analysis \cite{Carena:2002ss, Carena:2008vj}, however such contributions typically require larger values of $\phi_{\mu}$ to produce results consistent with the observed BAU and scale as $1/\tan \beta$ for large values; we do not consider such sources here.   Note that for large $\tan \beta$ one may be forced to consider larger values for $m_A$ to satisfy experimental bounds on the $B_u\rightarrow \tau \nu_{\tau}$ decay mode \cite{EWB_and_EDMs}.       

\subsection{Dependence on Bubble Wall Parameters}\label{sec:wall_param}
The baryon asymmetry also depends on several parameters associated with the electroweak phase transition, namely the bubble wall velocity, $v_w$, the wall width, $L_w$, and the variation of the ratio of Higgs vevs across the bubble wall, $\Delta\beta$.  An accurate determination of these quantities for a given choice of MSSM parameters requires solving for the exact Higgs profile in the bubble wall.  We instead aim to understand how uncertainties associated with the bubble wall parameters affect our results for the baryon asymmetry.  In this section we assess the various dependencies both qualitatively, by considering the analytic (approximate) solution to the QTEs discussed in detail in Refs.~\cite{Resonant_Relaxation, Chung:2008aya, Lepton_Mediated, Supergauge, Including_Yukawa}, and quantitatively by numerically solving the relevant set of QTEs while varying the relevant parameter values.  We find that our results are largely insensitive to $L_w$, mildly dependent on $v_w$, and linear in $\Delta\beta$.  We utilize these results to choose reasonable and, where appropriate, conservative values for the bubble wall parameters, leading to a more optimistic scenario for our results in the gaugino-higgsino mass planes, given our other assumptions about the MSSM parameter space.  

The baryon density $\rho_B$ satisfies a diffusion equation with solutions in the broken phase given in terms of $n_L(\bar{z})$ in the symmetric phase ($\bar{z}<0$) \cite{Resonant_Relaxation}, where $\bar{z}$ is the comoving distance away from the wall\footnote{In what follows we neglect the curvature of the wall}.  Early discussions (see e.g. \cite{Huet:1995sh, Resonant_Relaxation, More_Relaxed, CPV_and_EWB}) estimated the Yukawa rate, $\Gamma_Y$ (and strong sphaleron rate $\Gamma_{ss}$) to be much faster than all other relevant particle number-changing rates, implying that the higgsino density in the unbroken phase is quickly converted to a chiral quark density in front of the bubble wall.  This approximation allows one to write $Q,T$ in terms of $H$ and subsequently solve the QTEs by expanding in $1/\Gamma_{Y,ss}$ and rewriting the equations in terms of a single diffusion equation for $H$  \cite{Huet:1995sh} \begin{equation} \label{eq: diffusion}v_w H^{\prime} - \bar{D}H^{\prime \prime} = -\bar{\Gamma}H +\bar{S} \end{equation}  Here  $\bar{D}$ is a diffusion coefficient, $\bar{\Gamma}$ is a chiral relaxation term involving $\Gamma_Y$ as well as the CP--conserving stop mixing and higgsino-gaugino mixing rates, $\Gamma_M$, $\Gamma_H$, and statistical factors, $\bar{S}$ is a term proportional to the CP-violating sources, and primes denote derivatives with respect to $\bar{z}$.  Under the simplifying assumptions of step function CP-violating sources constant in the wall, and step function chiral relaxation rates, $\bar{\Gamma}(\bar{z})=\bar{\Gamma}\Theta(\bar{z})$, active only in the wall and broken phase, one obtains the lowest order solution for $H$ in the symmetric phase, \begin{gather} \label{eq:H_0} H_0(\bar{z}<0)=\frac{e^{v_w \bar{z}/\bar{D}}}{\bar{D}\kappa_+^2}\left(1-e^{-\kappa_+L_w}\right)\bar{S} \\ \label{eq:kap} \kappa_{\pm}=\frac{v_w\pm\sqrt{v_w^2+4\bar{D}\bar{\Gamma}}}{2\bar{D}} \end{gather}  Using Eq.(\ref{eq:H_0}) and the relations between $H$ and $Q,T$, one can obtain an analytic approximation for the baryon density in the broken phase to lowest order in $1/\Gamma_{y,ss}$ \cite{Resonant_Relaxation}.  

However, it was pointed out in Ref.~\cite{Including_Yukawa} that although the approximation of fast Yukawa rates will generally be valid in the symmetric phase, it does not typically apply in the broken phase.  Solving the diffusion equation to first non-trivial order in $1/\Gamma_{y,ss}$ yields corrections $\bar{S}\rightarrow\bar{S}+\delta\bar{S}+\mathcal{O}(1/\Gamma^{2}_{y})$ to the generalized source term in Eq. (\ref{eq: diffusion}).  Additionally, the expressions for the densities $Q$ and $T$, which to lowest order are proportional to $H$, receive corrections of $\mathcal{O}(1/\Gamma_{Y,ss})$, necessitating a full numerical solution to the set of coupled QTEs.  Despite these setbacks to reliably calculating the baryon asymmetry analytically, some insight regarding the dependence of $\rho_B$ on the various parameters can still be gained from the lowest order analytic solution and the first order corrections.

From Eqs.(\ref{eq:H_0})-(\ref{eq:kap}), we see that the baryon density depends nontrivially on the bubble wall velocity.  The largest corrections to the lowest order solution $H_0$ are typically those arising from the shift in the effective source term \cite{Including_Yukawa}, which can be written in terms of integrals over $H_0$ in the bubble wall.   These contributions depend on the wall velocity only through the combinations $\kappa_{\pm}$, which, in parametric regions where the corrections are large, depend only weakly on $v_w$.  This is because when the corrections $\delta\bar{S}/\bar{S}\sim\Gamma_M(\Gamma_H/\Gamma_Y)$ become large, $\bar{\Gamma}\sim\Gamma_M+\Gamma_H$ tends to dominate over $v_w^2$ in Eq. (\ref{eq:kap}).  Thus the velocity dependence of the baryon asymmetry can be reasonably approximated as that of the lowest order solution \begin{equation}\label{eq: approx} \rho_B(\bar{z}>0)\sim \frac{v_w \Delta\beta\bar{D}\mathcal{S}}{L_w\lambda_+\kappa_+^2(v_w -\bar{D}\lambda_-)}\left(1-e^{-\kappa_+L_w}\right)\left(\alpha_1+\alpha_2v_w^2 \right) \end{equation} where we have defined \begin{equation} \label{eq:lam} \lambda_{\pm}=\frac{v_w\pm \sqrt{v_w^2+4\mathcal{R}D_q}}{2D_q}, \end{equation} $D_q$ is an effective quark diffusion constant, $\mathcal{R}$ is a relaxation term arising from weak sphaleron processes in the unbroken phase, $\alpha_{1,2}$ are known functions independent of $v_w$, $L_w$, and $\Delta\beta$, and $\mathcal{S}$ is the effective source term after scaling out the dependence on the bubble wall parameters.  We have verified this approximate expression against the velocity dependence of the full numerical solution for various regions of the parameter space, both on and off resonance, and find the two to be in good agreement.  The profile Eq.(\ref{eq: approx}) reproduces well-known features of the velocity dependence of $\rho_B$, namely that small velocities correspond to a quasi-equilibrium situation, thereby suppressing the baryon asymmetry, while large velocities render the transport of the chiral current in front of the bubble wall inefficient, also suppressing the asymmetry, leading to a peak in $\rho_B$ for $v_w$ around a few $\times10^{-2}$ \cite{Huber:2001xf}. 

A precise determination of the bubble wall velocity in the MSSM is generally difficult, as the dynamics of the wall are further complicated by friction terms arising from the interactions of the wall with the plasma.  Detailed calculations of $v_w$ in the MSSM, including various frictional contributions, have been carried out \cite{vw}, suggesting a wall velocity in the range $10^{-2}<v_w<10^{-1}$.  To quantify the impact of this uncertainty on our results, we solve the QTEs numerically for wall velocities that maximize (minimize) the lowest order analytic approximation for $\rho_B$, Eq. (\ref{eq: approx}), at each point in the $(M_{1,2},\mu)$ plane.  The results are shown in Fig.\ref{fig:tb40_ewb}, with the band of maximal (minimal) velocities displayed in the inset.  We find that uncertainties in the wall velocity in this range lead to $\mathcal{O}(10$ GeV) uncertainties in the gaugino-higgsino mass plane which in turn do not significantly affect our conclusions.  In what follows we thus adopt the central value $v_w=.05$ except where otherwise stated.

Turning our attention to the dependence of the baryon asymmetry on the bubble wall width, $L_w$, we can once again look to Eq.(\ref{eq: approx}) for insight.  To lowest order in the Yuakwa and strong sphaleron rates, the baryon asymmetry is a monotonically decreasing function of $L_w$.  Corrections to Eq.(\ref{eq: approx}) in general do depend on $L_w$, however we find that including these corrections still renders $\rho_B$ a monotonically decreasing function of the wall width in our approximation and have verified this dependence for several choices of the relevant MSSM parameter values.  This behavior matches that expected of the sources, as, to lowest non-vanishing order in the Higgs vev insertion expansion \cite{Resonant_Relaxation}, the CP-violating higgsino source is proportional to the first spatial derivative of the Higgs vev in the wall and thus the baryon asymmetry becomes suppressed for larger values of $L_w$ \cite{Cline:2000kb, Huber:2001xf}.  In the analysis that follows we adopt the central value \cite{MSSM_Bubbles} $L_w=25/T$.  We find the impact of considering a much thinner bubble wall, $L_w \sim 5/T$, to be of $\mathcal{O}(1-10$ GeV$)$ in the $M_{1,2},\mu$ plane and therefore not substantially affecting our results.

We also see from Eq.(\ref{eq: approx}) that the baryon asymmetry is proportional to $\Delta\beta$, which is in turn a decreasing function of the CP-odd Higgs mass $m_A$.  For the $m_A$ dependence of $\Delta\beta$ we use the two-loop results of Ref.~\cite{MSSM_Bubbles} and obtain $\Delta\beta\sim4.5\times10^{-3}$ for $m_A=300$ GeV.  As mentioned above, we find the baryon asymmetry to be significantly suppressed for smaller values of $\Delta\beta$ (corresponding to larger $m_A$; see e.g. Fig.\ref{fig:tb40_2}) and we thus take this choice to represent a conservative bound on the $\Delta\beta$ dependence of $Y_B$, although in some cases we do consider the $m_A=1000$ GeV scenario for illustrative purposes.  The $\tan\beta$ dependence of $\Delta\beta$ has not, to our knowledge, been thoroughly explored in the literature \cite{EWB_and_EDMs}, however from Eq.(\ref{eq: approx}) its effect on $Y_B$ would seem to simply rescale our curves by a constant multiplicative factor for a given choice for $\tan\beta$.  As discussed in Sec.~\ref{sec:param}, we do not anticipate the inclusion of such effects to greatly impact our conclusions and we defer such considerations to future work.

Finally, we note that the suppression of the net baryon density for large values of $m_{\widetilde{t}_R}$ mentioned in the previous subsection is manifest in the approximate solution for the baryon density Eq.(\ref{eq: approx}).  The $\alpha_{1,2}$ depend on the RH stop mass through the statistical factor $k_T$, which relates the RH (s)top chemical potential to the corresponding number density.   For increasing $m_{\widetilde{t}_R}$, the stop contribution to $k_T$ becomes exponentially small and $\alpha_{1,2}$ decrease to their asymptotic values for large $m_{\widetilde{t}_R}$ (with all other relevant parameters fixed), while the $\alpha_2$ term (which is generally positive) is proportional to $1/\Gamma_{ss}$, resulting in a suppressed baryon density for realistic strong sphaleron rates when compared to the $\Gamma_{ss}\rightarrow 0$ case for large $m_{\widetilde{t}_R}$ \cite{Giudice:1993bb,Huet:1995sh}.

\section{EWB and Bino-like Dark Matter}\label{sec:bino}

\begin{figure*}[!t]
\mbox{\hspace*{-1.2cm}\includegraphics[width=0.55\textwidth,clip]{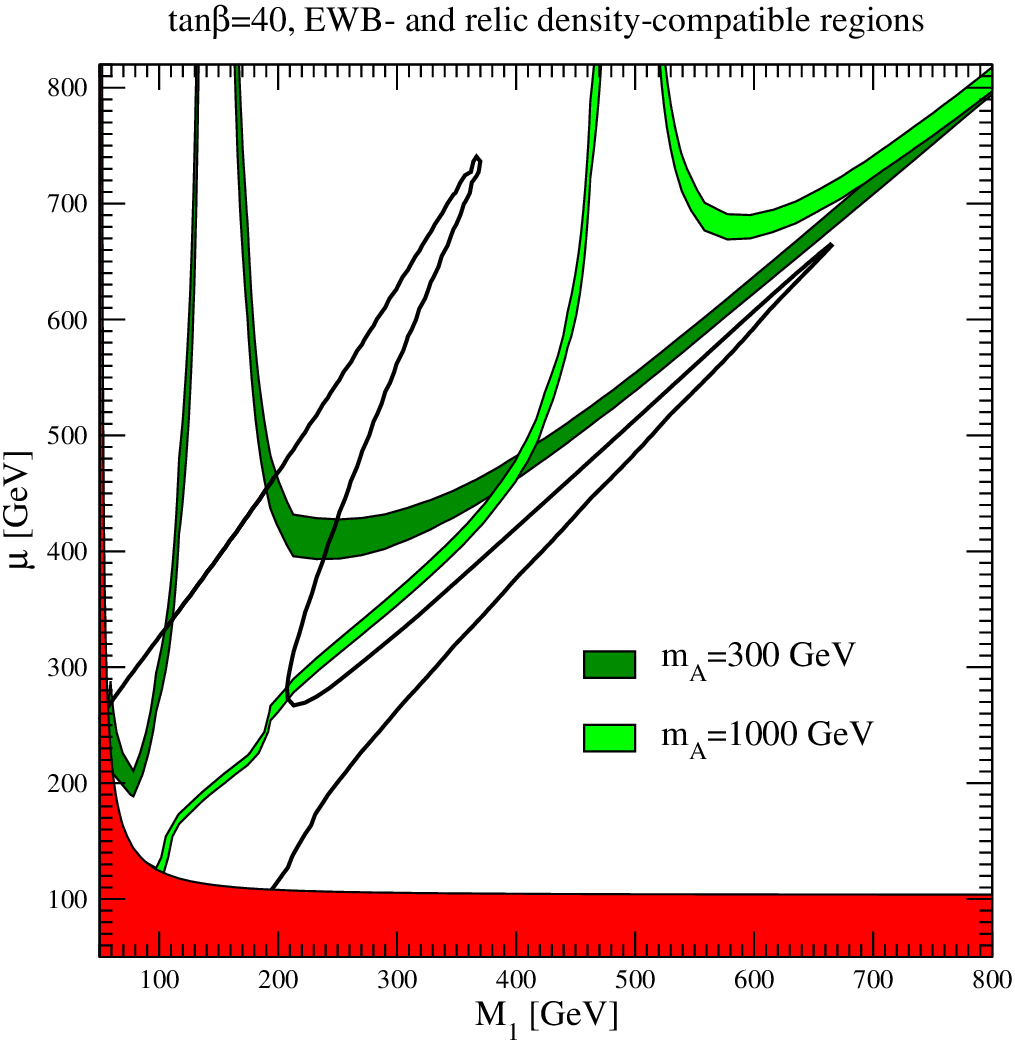}\qquad\includegraphics[width=0.55\textwidth,clip]{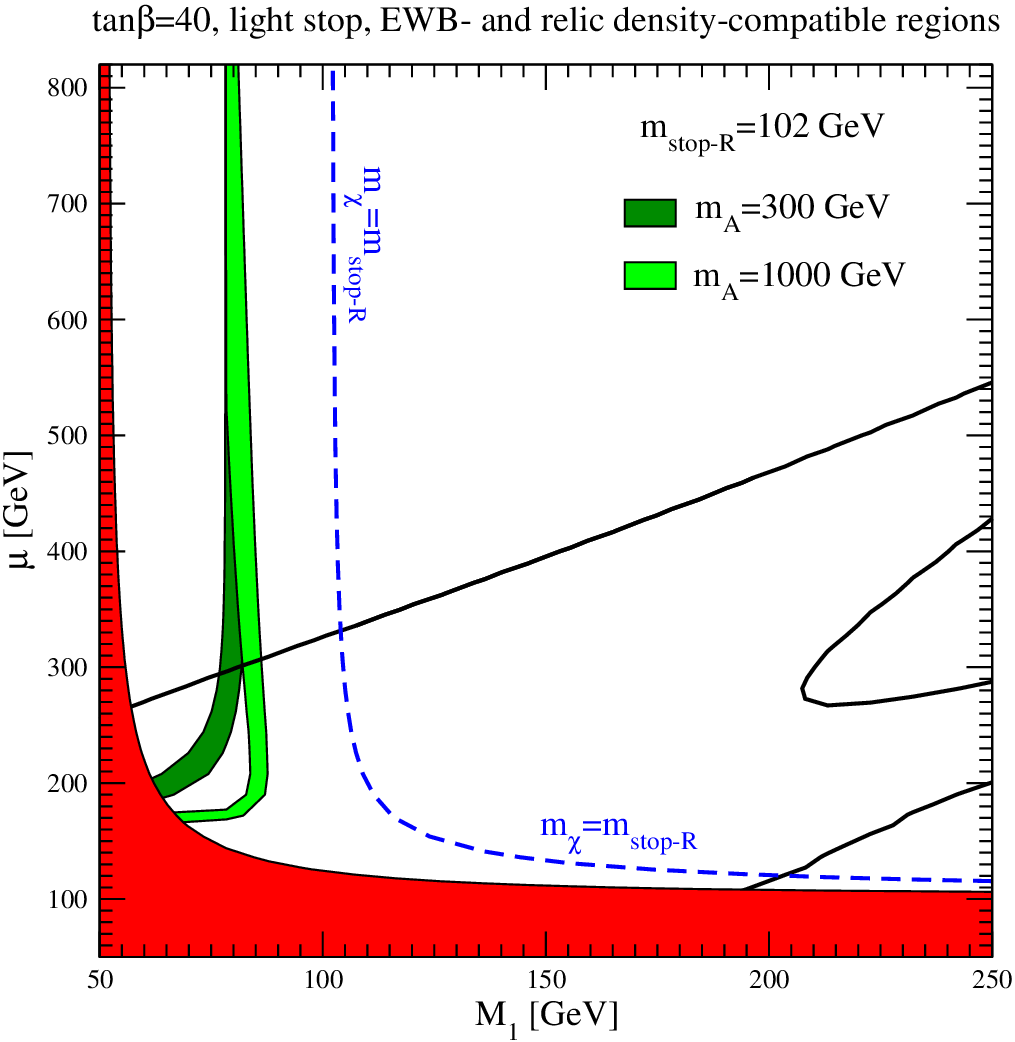}}
\caption{\label{fig:tb40_1}\it\small Left: Regions with a thermal relic neutralino abundance $0.095<\Omega_\chi h^2<0.13$ on the $(M_1,\mu)$ plane at $\tan\beta=40$, for maximal gaugino-higgsino CP-violating phase $\sin\phi_\mu=1$ and for $m_A=300$ GeV (dark green) and $m_A=1000$ GeV (light green). The black line indicates successful electroweak baryogenesis for $\sin\phi_\mu=1$ and $m_A=300$ GeV, as in Fig.~\ref{fig:tb40_ewb}. Right: Regions of correct relic abundance, with a lightest stop set at a mass of 102 GeV, again for $m_A=300$ and 1000 GeV (darker and lighter green, respectively). The dashed blue line corresponds to the parameter space where the lightest neutralino has the same mass as the lightest stop. Bounds on the density of heavy relic charged or colored particles exclude the portion of the parameter space above and to the right of the dashed blue line, where the stop would be the lightest supersymmetric particle.}
\end{figure*}

Having outlined the portions of the MSSM parameter space relevant for successful resonant EWB, we now direct our attention to the dark matter phenomenology associated with these regions.  As mentioned in Sec.~\ref{sec:asymmetry}, the lightest neutralino in successful supersymmetric EWB models is, potentially, a viable dark matter candidate. We also pointed out above that the phenomenology of the lightest neutralino depends sensitively on the hierarchy between the masses of the hypercharge and weak gauginos, which is not fixed by EWB. In practice, EWB dictates that either $M_1\sim\mu$ or $M_2\sim\mu$, but it does not enforce a hierarchy between $M_1$ and $M_2$ which depends, in general, on the mechanism of supersymmetry breaking. We first consider the case where $M_1<M_2$, which occurs for example in models where gaugino soft breaking masses are universal at the grand unification scale. Renormalization group evolution then dictates, approximately, that $M_2\simeq\frac{3}{5}\frac{\cos^2\theta_W}{\sin^2\theta_W}M_1\simeq2M_1$.

Having established a relation between the soft-breaking gaugino masses, the mass and composition of the lightest neutralino only depends on the values of $M_1$ and $\mu$. We begin exploring the dark matter phenomenology on the $(M_1,\mu)$ parameter space in Fig.~\ref{fig:tb40_1}. In the left panel, we calculate the relic density and show regions on the parameter space where the thermal relic density of the lightest neutralino $\Omega_\chi$ falls in a range consistent with the inferred dark matter density in the universe \cite{Komatsu:2010fb} (quantitatively, we highlight regions of parameter space where $0.095<\Omega_\chi h^2<0.13$). We fix the mass scale of the heavy MSSM Higgs sector by setting $m_A=300$ GeV and $m_A=1000$ GeV - two values that bound the interesting range for MSSM EWB as discussed in Sec.~\ref{sec:param}.

The shape of the regions where the lightest neutralino relic density matches the observed density of dark matter are qualitatively easily understood: to produce a large enough abundance, interactions of neutralinos with gauge bosons must be suppressed, enforcing a bino-like character to neutralino relics with the correct abundance; the two vertical funnels then correspond to rapid, quasi-resonant annihilation via the CP-odd Higgs $A$ when $m_\chi\simeq m_A/2$, while away from the resonance the bino relic density is low enough only if a sufficiently large higgsino fraction is present -- enforcing $M_1\simeq\mu$. Notice that enforcing successful EWB as well as the correct relic abundance implies a bino-like neutralino and $M_2\sim\mu$ to produce enough baryon asymmetry. The $M_1\sim \mu$ funnel of neutralino-driven EWB \cite{Li:2008ez} lies not far from, but well below, the parameter space with the correct thermal relic abundance of neutralinos. Notice also that the regions with the desired overlap of relic density and EWB depend upon the choice of the CP-violating phase, which for the black lines shown in the figure is maximal.

In the context of the minimal supersymmetric extension of the Standard Model, the electroweak phase transition is strongly first order and compatible with collider data only for a very light right-handed stop and in a certain mass window for the Higgs mass. Ref.~\cite{Carena:2008vj} most recently addressed  this issue, in the context of an effective theory with decoupled sfermion (with the exception of the right-handed stop) and heavy-Higgs sectors \cite{Carena:2008rt}. The allowed region is restricted to right-handed stop masses lighter than around 115 GeV. Such a light stop has dramatic implications not only for collider phenomenology (see e.g. \cite{Bornhauser:2010mw} and references therein), but also for dark matter searches: the lightest neutralino must be lighter than the stop in order for the model to be viable. As a result, the range of neutralino masses is severely restricted. In addition, stop coannihilation also occurs \cite{Balazs:2004ae}, when the masses of the lightest neutralino and stop approach each other, and the freeze-out of the two species in the early universe is correlated. 

To illustrate this point, we outline on the right in Fig.~\ref{fig:tb40_1} the regions in the $(M_1,\mu)$ parameter space compatible with a neutralino thermal relic abundance matching the cold dark matter density, for $m_A=300$ and 1000 GeV (darker and lighter green, respectively) when a right-handed stop with a mass of 102 GeV is assumed. To the right of the vertical allowed bands, the relic density is driven to excessively low values via the mechanism of stop coannihilation, while for low values of $M_1$ coannihilations are ineffective and the bino relic density is too large. The plot also shows the boundary of the ``allowed region'' where the lightest neutralino is the LSP (dashed blue line), and the fact that there is an overlap between the correct relic density regions and the regions with enough baryon asymmetry (black line) \footnote{Note that other non-resonant CP-violating sources not considered here might also contribute to the BAU in this scenario, especially for larger values of $m_A$ \cite{Carena:2008vj}}. In the remainder of this analysis, we omit the curves corresponding to the light stop scenario, but the Reader should bear in mind that the shape indicated by the dashed blue line would appear in an analogous way in all other figures, should one resort to the light-stop minimal scenario.

\begin{figure*}[!t]
\mbox{\hspace*{-1.2cm}\includegraphics[width=0.55\textwidth,clip]{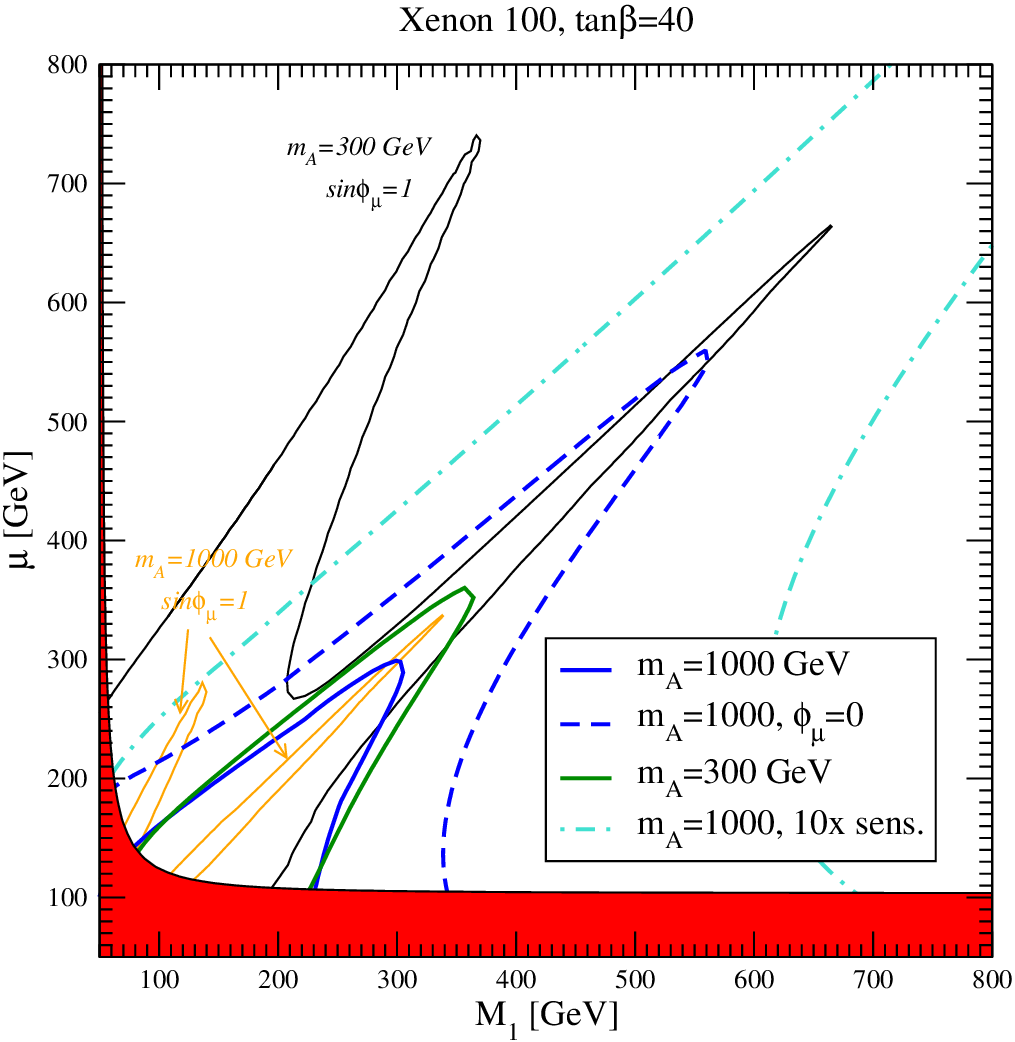}}
\caption{\label{fig:dirdet_1}\it\small Regions excluded by the Xenon100 direct detection results, on the $(M_1,\mu)$ plane at $\tan\beta=40$, for maximal gaugino-higgsino CP-violating phase $\sin\phi_\mu=1$ and for $m_A=1000$ GeV (solid blue line) and 300 GeV (green line). The dashed blue line corresponds to no CP violation, and $m_A=1000$ GeV. Finally, the turquoise dot-dashed line indicates the region corresponding to 10 times the Xenon100 current sensitivity. The black solid line outlines the region of parameter space where successful EWB can occur, for $m_A=300$ GeV, while the orange line corresponds to $m_A=1000$ GeV.}
\end{figure*}

Turning our attention to the various dark matter searches, we outline the impact of the recent direct detection results from the Xenon 100 collaboration \cite{Aprile:2011hi} on the parameter space in Fig.~\ref{fig:dirdet_1} . For reference, we show the contours corresponding to successful EWB for both $m_A=300$ GeV and 1000 GeV (orange solid lines). The Xenon 100 constraints are calculated by computing for every point in the parameter space the spin-independent neutralino-proton cross section and comparing with the cross section limit corresponding to the given WIMP mass. As observed in Ref.~\cite{EWB_and_DM}, the relevant Higgs-neutralino coupling is sensitive to the relative bino-higgsino CP-violating phase $\phi_\mu$, with larger phases suppressing the spin-independent neutralino-proton cross section. 

We show with a blue solid line the contours of the region excluded by the results from the Xenon 100 collaboration presented in Ref.~\cite{Aprile:2011hi} for $m_A=1000$ GeV and $\sin\phi_\mu=1$. Points between the red regions and the contours are excluded. Notice that almost the entire $M_1\sim\mu$ funnel of successful EWB is ruled out by the Xenon 100 results. Turning off CP violation extends the contours significantly, to the blue dashed lines, indicating that for smaller CP-violating phases the $M_1\sim\mu$ funnel is solidly excluded by direct detection. The green solid line encompasses the slightly larger region corresponding to $m_A=300$ GeV: a lower value for $m_A$ produces larger regions compatible with EWB, and the $M_1\sim\mu$ funnel, for large enough CP violation, is still a viable option for successful EWB. Finally, for reference we show the exclusion limits that would correspond to an improvement in the Xenon 100 sensitivity by a factor of 10. Such an improvement in sensitivity is likely optimistic as a target for Xenon 100 by the end of 2012, but is well within the sensitivity expected for the XENON1T experiment, recently approved by INFN to start at the Laboratori Nazionali del Gran Sasso \cite{elenaquote}, even with a very limited time exposure. A gain in sensitivity of a factor 10 would vastly probe the entire region of $M_1\sim\mu$ as well as the low-mass portion of the $M_2\sim\mu$ funnel.

\begin{figure*}[!t]
\mbox{\hspace*{-1.2cm}\includegraphics[width=0.55\textwidth,clip]{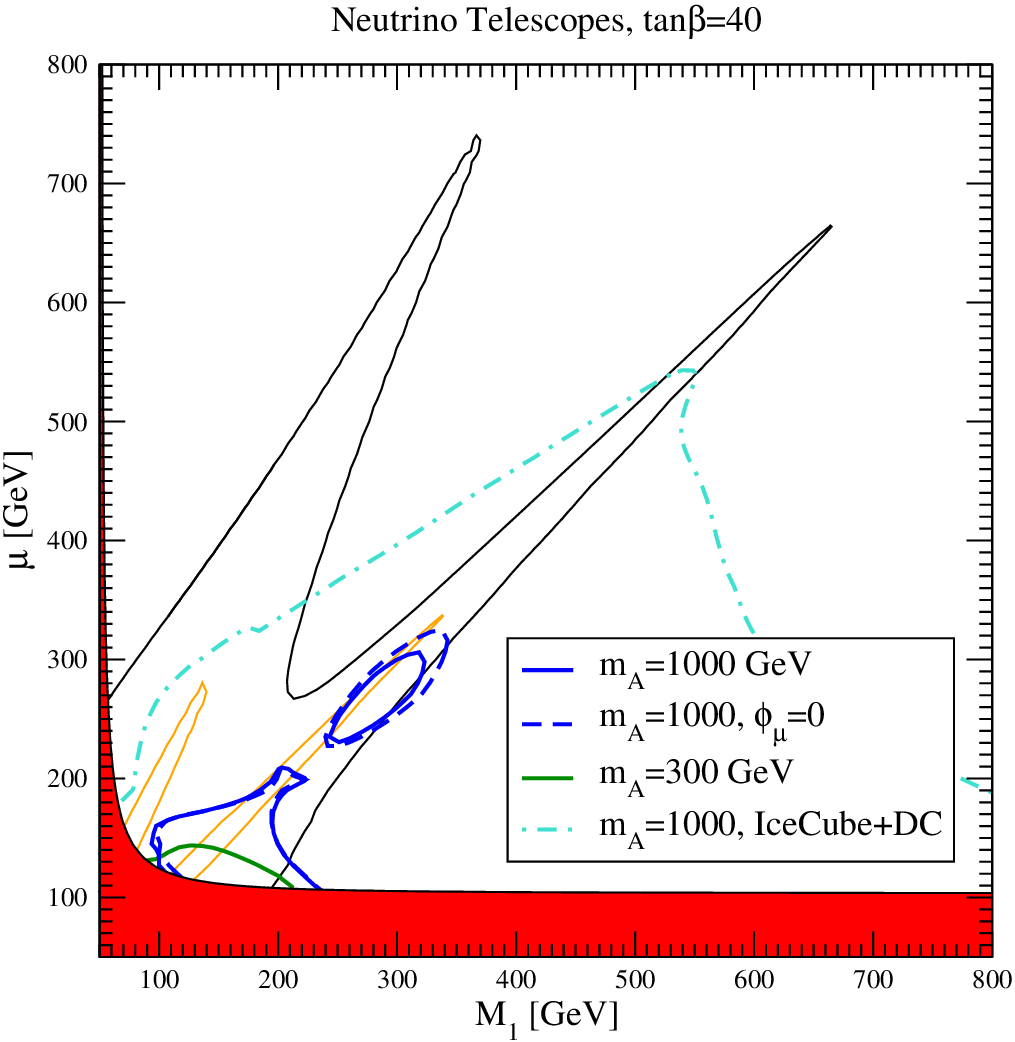}\qquad\includegraphics[width=0.55\textwidth,clip]{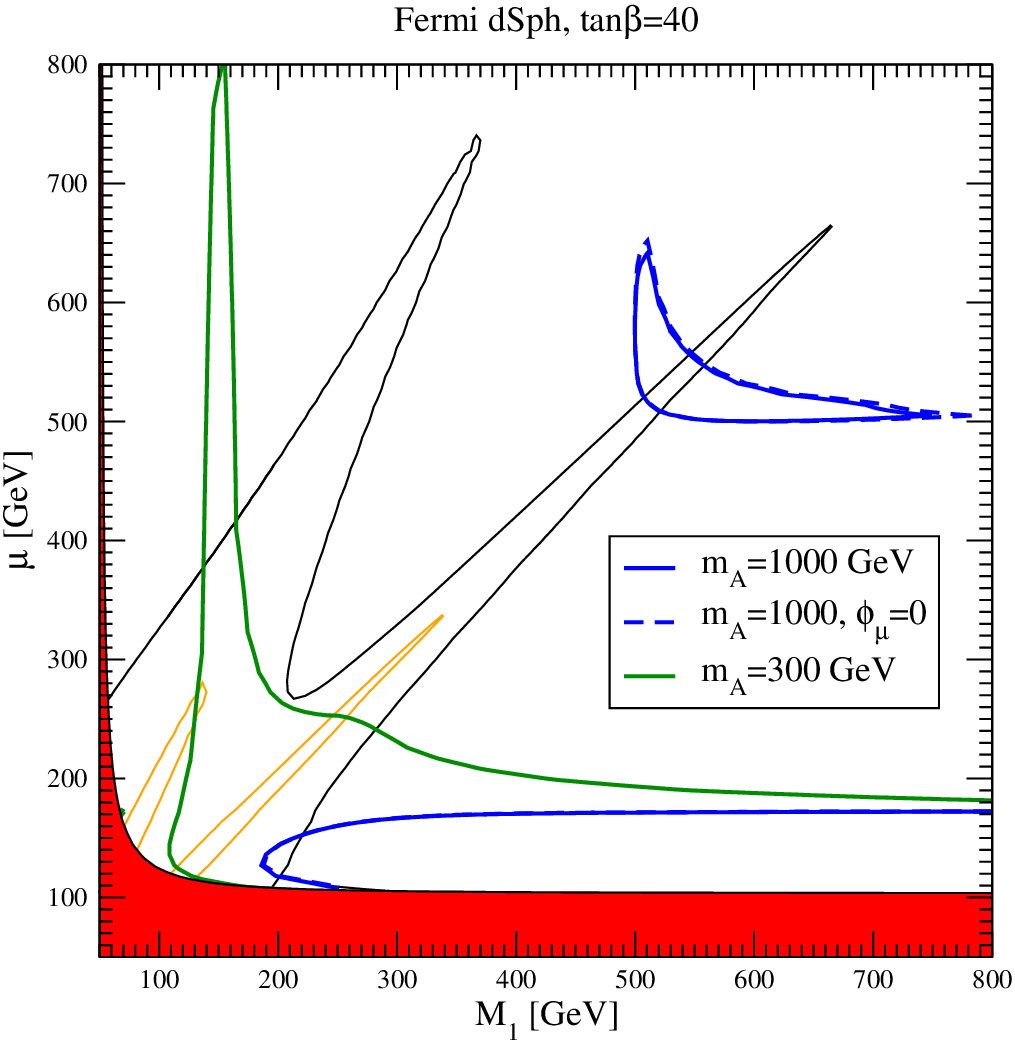}}
\caption{\label{fig:tb40_2}\it\small The performance of indirect dark matter searches on the $(M_1,\mu)$ plane. In the left panel, we show the current and future reach of the IceCube neutrino telescope \cite{icecube}, while on the right we indicate the regions of parameter space ruled out by Fermi observations of nearby dwarf spheroidal galaxies \cite{dsph}}
\end{figure*}

We move on to indirect dark matter detection in Fig.~\ref{fig:tb40_2}. The left panel shows the current performance of the IceCube detector in searching for an anomalous high-energy flux of neutrinos from the Sun that might originate from the annihilation of dark matter captured and sank inside the Sun \cite{icecube}. Neutrino telescope rates are sensitive primarily to the capture rate, which depends upon the vector and scalar neutralino-proton scattering cross section. In turn, the latter is maximized for maximal higgsino-gaugino mixing in the lightest neutralino. This is manifest in the shape of the parameter space regions ruled out by current IceCube data \cite{icecube}, primarily covering the $M_1\sim\mu$ funnel, albeit only for relatively low masses - at most $m_\chi\lesssim300$ GeV. We also show the effect of switching off the CP-violating phase $\phi_\mu$, which bears for neutrino telescope rates the opposite effect as for the spin-independent cross section (see again the detailed discussion in Ref.~\cite{EWB_and_DM}), here enhancing -- although only marginally -- the telescope sensitivity to the theory parameter space. This is shown by the dashed blue line. A lighter heavy-Higgs sector suppresses the reach of neutrino telescopes, as evidenced by the green line. This unusual effect is due to the bias in the annihilation final state produced by resonant annihilations via the $A$ CP-odd Higgs boson, that produces more $b\bar b$ than gauge boson pairs in the final state. In turn, this yields softer neutrinos, which are harder to detect with IceCube. Finally, the dot-dashed light-blue line shows the anticipated performance of the full IceCube instrument, including DeepCore, with 180 days of data \cite{icecube}. Again, large portions of the parameter space where the lightest neutralino is relatively light (less than 0.5 TeV) and with a large higgsino-bino mixing will be readily ruled out by forthcoming IceCube results.

The right panel of Fig.~\ref{fig:tb40_2} shows the impact of observations of dwarf spheroidal galaxies (dSph) with the Fermi gamma-ray telescope \cite{dsph}. The null result of searches for gamma-ray emission from dSph is translated into a limit on the pair-annihilation cross section, after including an appropriate normalization factor dependent upon the dark matter density distribution in the relevant dSph, and utilizing the appropriate gamma-ray spectrum (we refer the Reader to Ref.~\cite{dsph} for further details on this analysis). The parameter space regions that are excluded by Fermi data correspond to regions with large pair-annihilation cross section (we do not rescale here for under-abundant relic dark matter density, assuming that non-thermal production accounted for low-thermal relic abundances). Two regions are ruled out by Fermi data: in the lower right light higgsino-like neutralinos, and for $m_\chi\simeq m_A/2$ resonantly annihilating neutralinos.

We caution the Reader that the conclusions we arrive at in the present analysis crucially depend on the assumption that the dominant sources at the electroweak phase transition correspond to resonant chargino-neutralino terms. Should one allow for additional source terms, such as non-resonant sources (see e.g. \cite{Carena:2008vj}) or sources associated with left-right stops, sbottoms or staus (see e.g. \cite{Including_Yukawa}), all of our conclusions would be weakened. For example, non-resonant sources allow for large values of $m_A$ that would suppress the spin-independent dark matter direct detection rates. Losing the correlation between the $\mu$ parameter and the gaugino soft supersymmetry-breaking masses would also impact the higgsino mixing in the lightest neutralino, again affecting virtually all direct and indirect detection rates.

\section{EWB and Wino-like Dark Matter}\label{sec:wino}

\begin{figure*}[!t]
\mbox{\includegraphics[width=0.55\textwidth,clip]{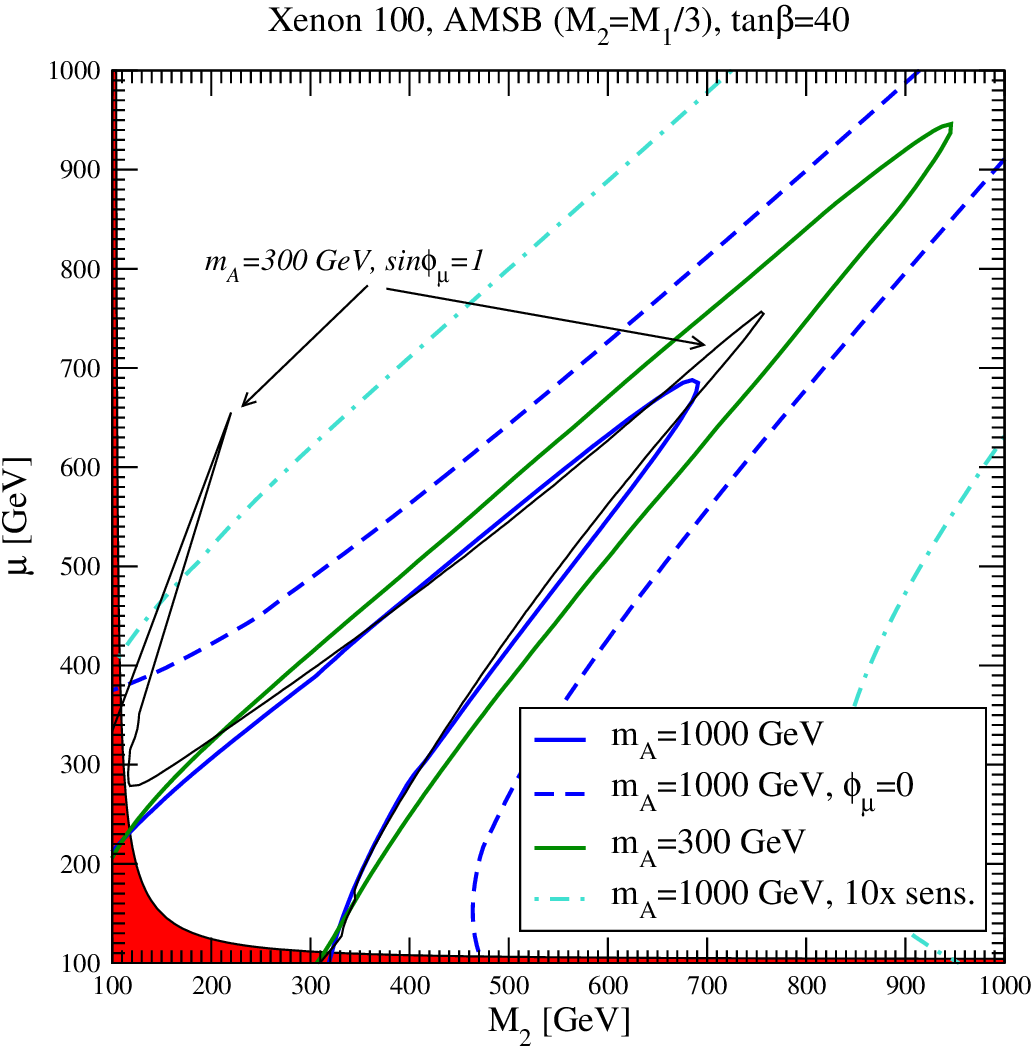}}
\caption{\label{fig:dirdet_amsb}\it\small As in Fig.~\ref{fig:tb40_1}, right, but on the $(M_2,\mu)$ plane, and with an anomaly-mediated gaugino mass hierarchy (whereby $M_2\simeq M_1/3$).}
\end{figure*}

If the hierarchy between the gaugino masses at the electroweak scale is such that $M_2<M_1$, the phenomenology of the lightest neutralino as a dark matter particle changes drastically from that just discussed. A prototypical scenario where the $M_2<M_1$ hierarchy is realized is anomaly mediated supersymmetry breaking (see e.g. \cite{amsb}). Although this class of models is unrelated to the phenomenological requirements of EWB, we employ here for definiteness the prediction, valid at the electroweak scale, that $M_2\simeq M_1/3$. The lightest neutralino is here either wino-like, higgsino-like, or a mixed wino-higgsino state, depending upon the relative size of $\mu$ versus $M_2$. For sub-TeV masses, in no case is the lightest neutralino thermal relic density large enough to produce the inferred density of cold dark matter. Requiring that the lightest neutralino be the dark matter constituent therefore necessitates one to postulate a non-thermal production mechanism \cite{amsb} or a modified cosmological setup for example with kination domination \cite{quintess}. Here, as before, we assume that the density of neutralinos matches the cold dark matter density by one of these mechanisms.
 
Fig.~\ref{fig:dirdet_amsb} illustrates, as in Fig.~\ref{fig:tb40_1}, the parameter space probed by the most recent Xenon 100 results on the spin-independent neutralino-proton scattering cross section, for a variety of choices for $m_A$ and $\phi_\mu$. The black lines indicate the maximal extent of the EWB-compatible parameter space, corresponding to maximal CP violation, i.e. $\sin\phi_\mu=1$, within the present setup. Current direct detection results (solid blue and green curves, for $m_A=1000$ and 300 GeV, respectively) rule out almost the entire EWB-compatible parameter space, even for the most unfavorable case of maximal CP violation. Increasing $\sin\phi_\mu$ leads to a reduced EWB-viable region and a wider portion of parameter space ruled out by the direct detection results, as illustrated by the blue dashed line. The only region of parameter space not constrained by direct searches for dark matter is that where $M_1\sim\mu$ (one should however bear in mind that the location of this sliver of the parameter space depends on the details of the gaugino mass hierarchy). An improvement by a factor of 10 in the direct detection sensitivity will significantly extend the parameter space excluded by dark matter searches, as illustrated by the dot-dashed light-blue line.

\begin{figure*}[!t]
\mbox{\hspace*{-1.2cm}\includegraphics[width=0.55\textwidth,clip]{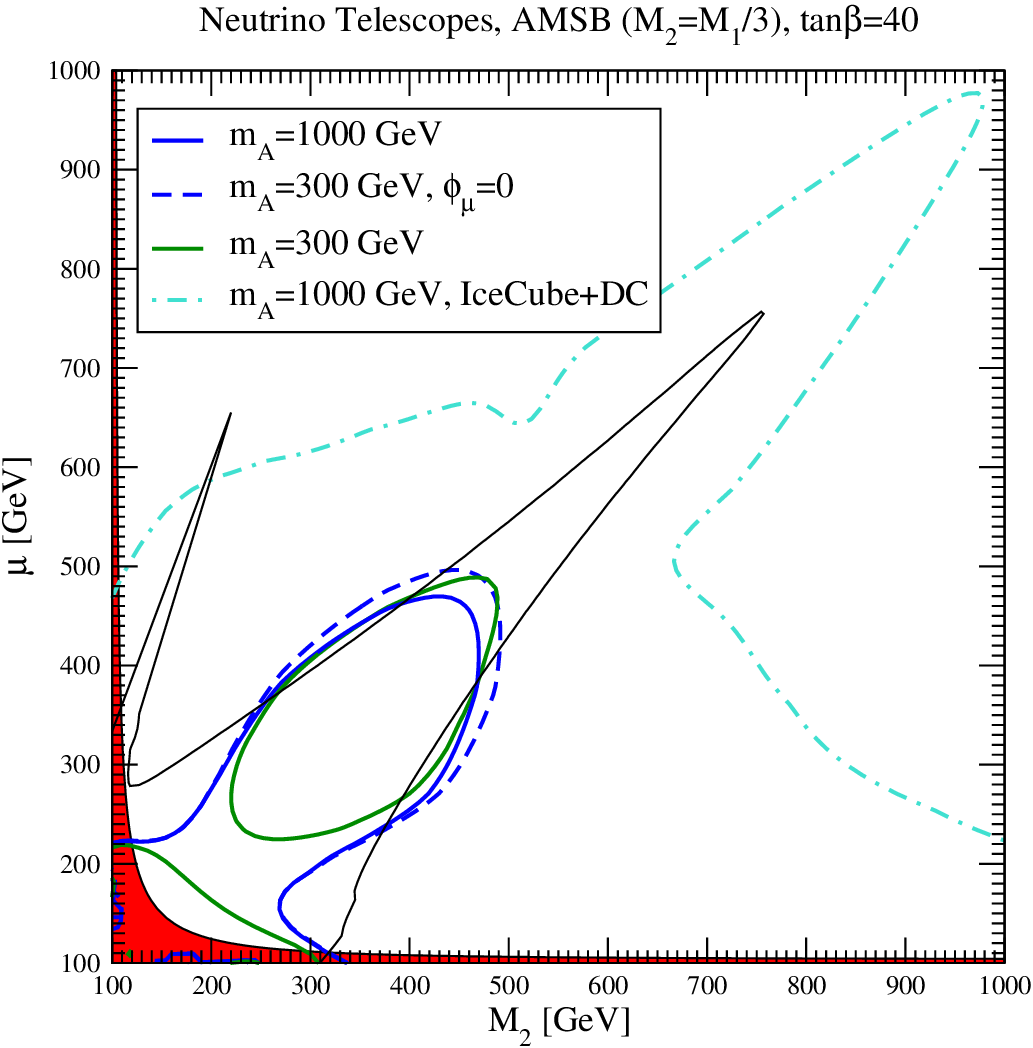}\qquad \includegraphics[width=0.55\textwidth,clip]{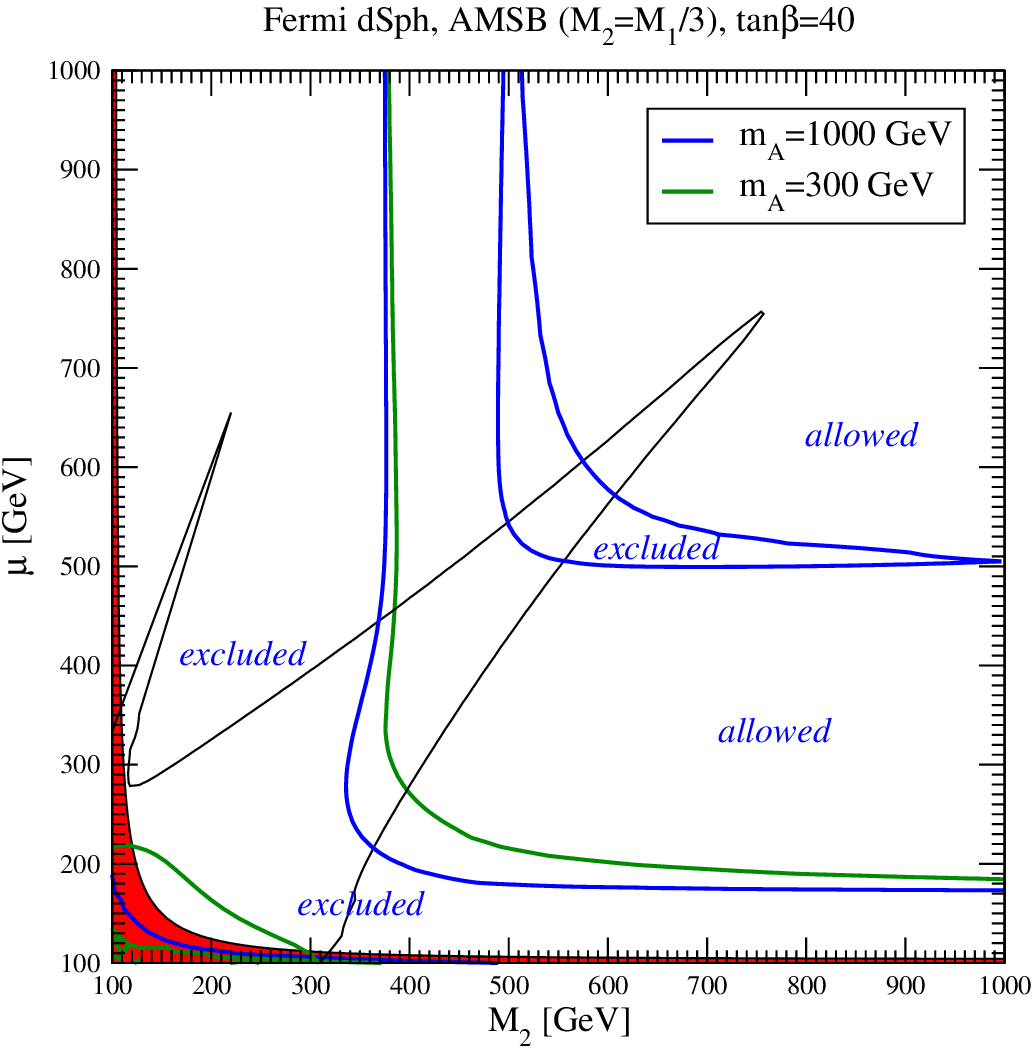}}
\caption{\label{fig:indirect_amsb}\it\small As in Fig.~\ref{fig:tb40_2}, but on the $(M_2,\mu)$ plane, and with an AMSB hierarchy (whereby $M_2\simeq M_1/3$).}
\end{figure*}

Neutrino telescopes also very effectively probe wino-like dark matter in the context of an MSSM realization with successful EWB, as shown by the left panel of Fig.~\ref{fig:indirect_amsb}. The effect of resonant neutralino annihilation via the CP-odd Higgs $A$ is shown clearly in the shape of the exclusion limits, including the the curve outlining the future performance of the full IceCube plus DeepCore detector. Current IceCube data \cite{icecube} probes neutralinos as heavy as almost 0.5 TeV, but leaves both the bino-driven EWB funnel and the high-mass tip of the wino-driven EWB funnel as viable parameter space regions.

Finally, the right panel of Fig.~\ref{fig:indirect_amsb} indicates the regions ruled out by Fermi observations of dSph. Here, similarly to the $M_1<M_2$ case, the parameter space regions excluded by gamma-ray data mainly depend upon the size of the ratio $\langle\sigma v\rangle/m_\chi^2$. The excluded regions correspond to wino or higgsino-like light neutralinos, or to resonantly annihilating heavier neutralinos with a mass $m_\chi\simeq m_A/2$. Remarkably, we find that the $M_1=3M_2\sim\mu$ funnel is entirely ruled out by gamma-ray data, as it falls in a region of light wino-like dark matter.

We thus find that MSSM EWB with an $M_2<M_1$ gaugino mass hierarchy is essentially ruled out by dark matter searches, if the lightest neutralino is the main dark matter constituent. Also, wino-like dark matter is not compatible with EWB, as illustrated by the upper left halves of Figs.~\ref{fig:dirdet_amsb}-\ref{fig:indirect_amsb}. An EWB model with $M_2<M_1$ would thus only be phenomenologically viable if either $R$-parity were violated, and the lightest neutralino were not stable, or if the lightest neutralino were not the dominant dark matter constituent and some other particle were the dark matter.

We emphasize that these results (and those in the previous sections) follow largely from the relative hierarchy of the gaugino masses and do not depend sensitively on their particular values.  On the one hand, the dark matter search constraints are determined primarily by the lightest gaugino mass and $\mu$ and are insensitive to the precise mass of the heavier gaugino.  On the other hand, regions compatible with successful EWB depend on the relation between $\mu$ and $M_{1,2}$; changing the details of the gaugino mass hierarchy affects the curves of constant $Y_B$ in Figs.~\ref{fig:tb40_ewb}-\ref{fig:indirect_amsb} by squeezing the upper funnel either towards the lower funnel or towards the $\mu$-axis.  If pushed downwards, larger portions of these EWB-compatible regions will typically be probed by direct detection and neutrino telescope data, which inherently constrain the lightest gaugino $\sim\mu$ funnel (this is where the LSP has maximal gaugino-higgsino mixing).  If the upper funnel is pushed towards the $\mu$-axis, portions of this region for the $M_1<M_2$ -type hierarchy will generally still be allowed, as observations of dSph will mostly probe higgsino-like and resonantly annihilating LSPs.  For the $M_2<M_1$ case, however, gamma ray telescopes will additionally constrain the upper left half of the $(M_2,\mu)$ plane where any potentially successful bino-diven EWB will take place (this funnel is always smaller in size than the $M_2\sim\mu$ funnel because the bino-higgsino resonance cannot occur through chargino exchange \cite{EWB_and_DM}), thus generally combining with the direct detection and neutrino telescope results to rule out virtually all of the parameter space viable for EWB, independent of the details of the gaugino masses. 

\section{EWB and the Large Hadron Collider}\label{sec:lhc}

\begin{figure*}[!t]
\mbox{\hspace*{-1.2cm}\includegraphics[width=0.55\textwidth,clip]{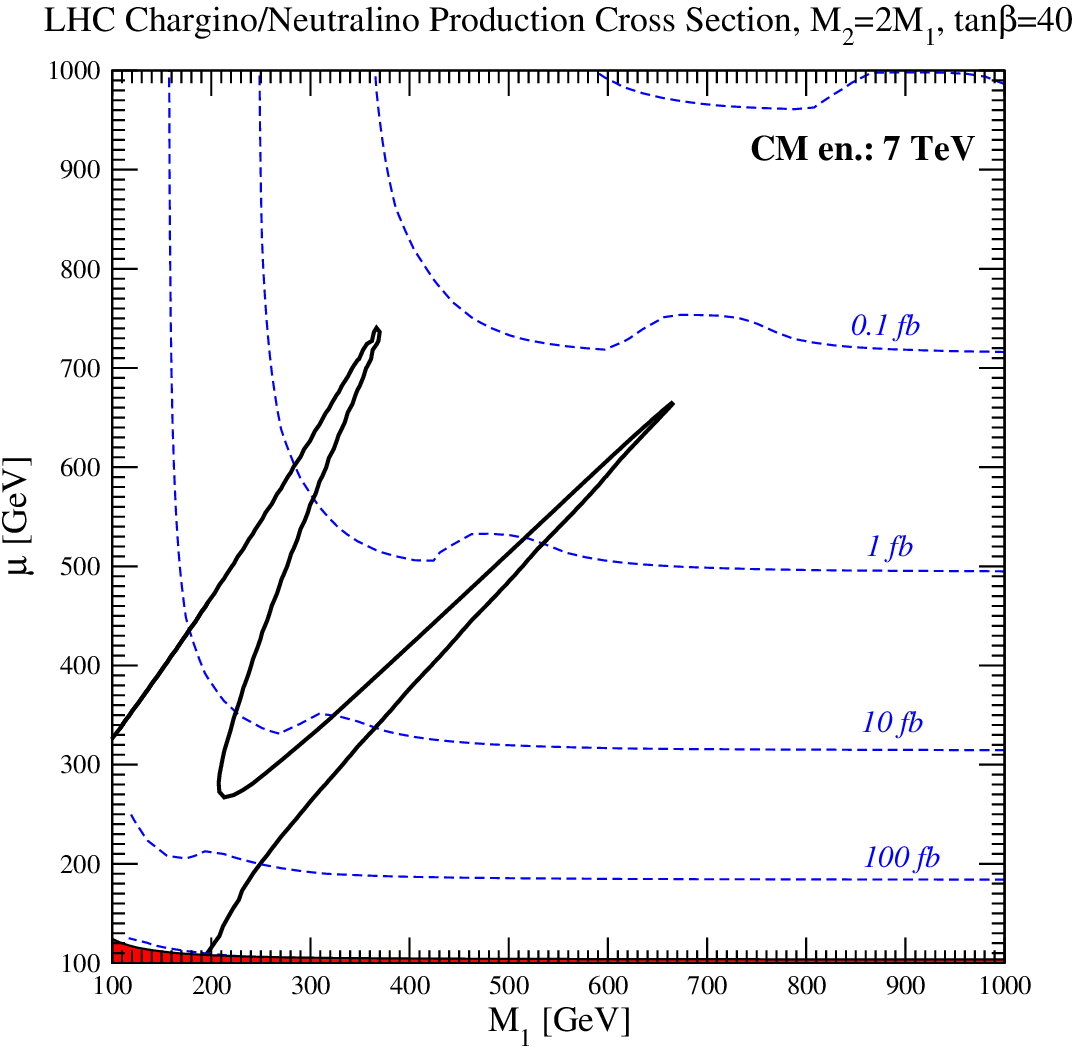}\qquad \includegraphics[width=0.55\textwidth,clip]{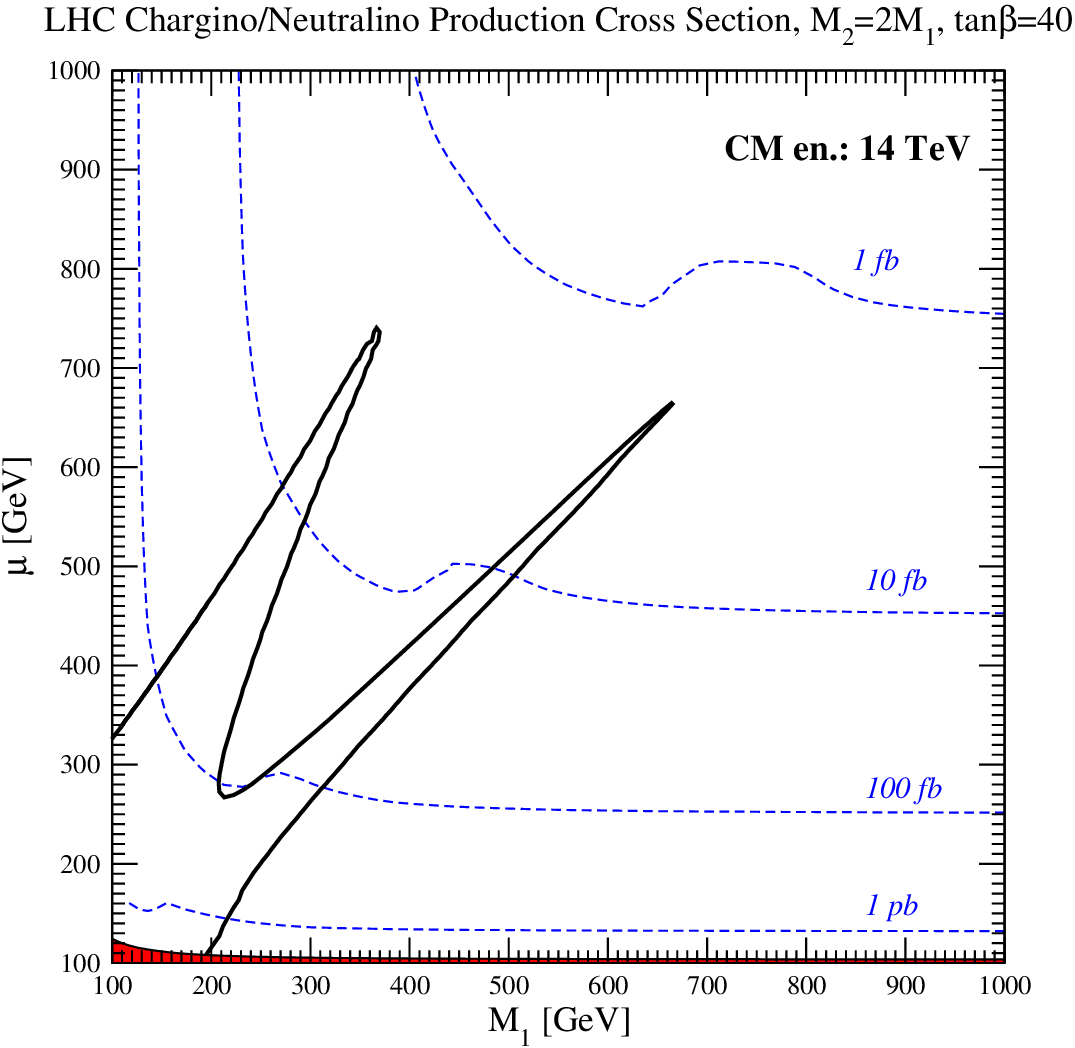}}
\caption{\label{fig:lhc_tb40}\it\small Curves of constant total -ino production cross-section at the LHC for points in the $(M_1,\mu)$ plane for the gravity-mediated supersymmetry breaking gaugino mass hierarchy, with $\sqrt{s}=7$ TeV (left) and 14 TeV (right), and with $\tan\beta=40$, $m_A=300$ GeV and all other parameters as discussed in the text.  Also shown are regions compatible with resonant chargino-neutralino electroweak baryogenesis, on the $(M_1,\mu)$ plane for the same values of the other parameters and maximal gaugino-higgsino CP-violating phase $\sin\phi_\mu=1$.}
\end{figure*}

\begin{figure*}[!t]
\mbox{\hspace*{-1.2cm}\includegraphics[width=0.55\textwidth,clip]{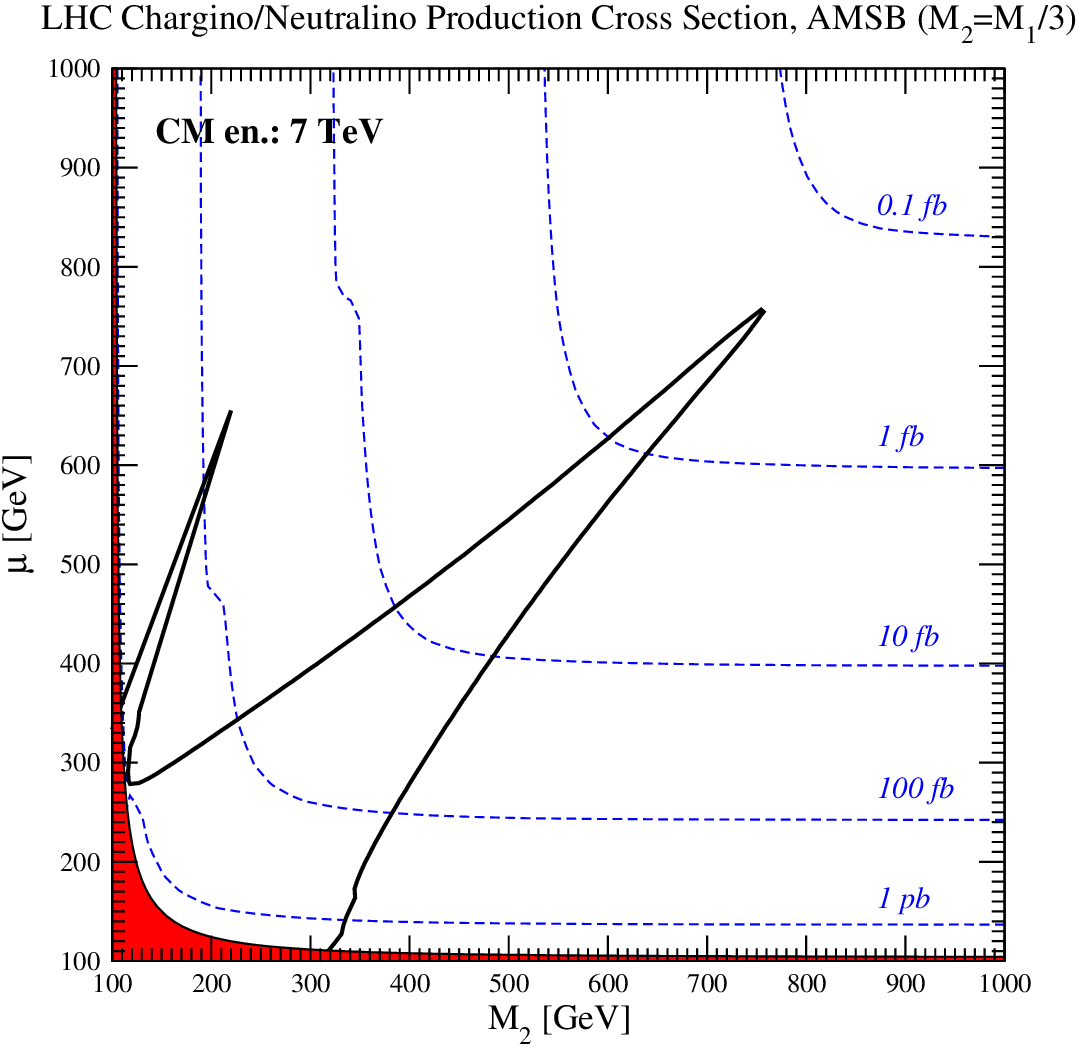}\qquad\includegraphics[width=0.55\textwidth,clip]{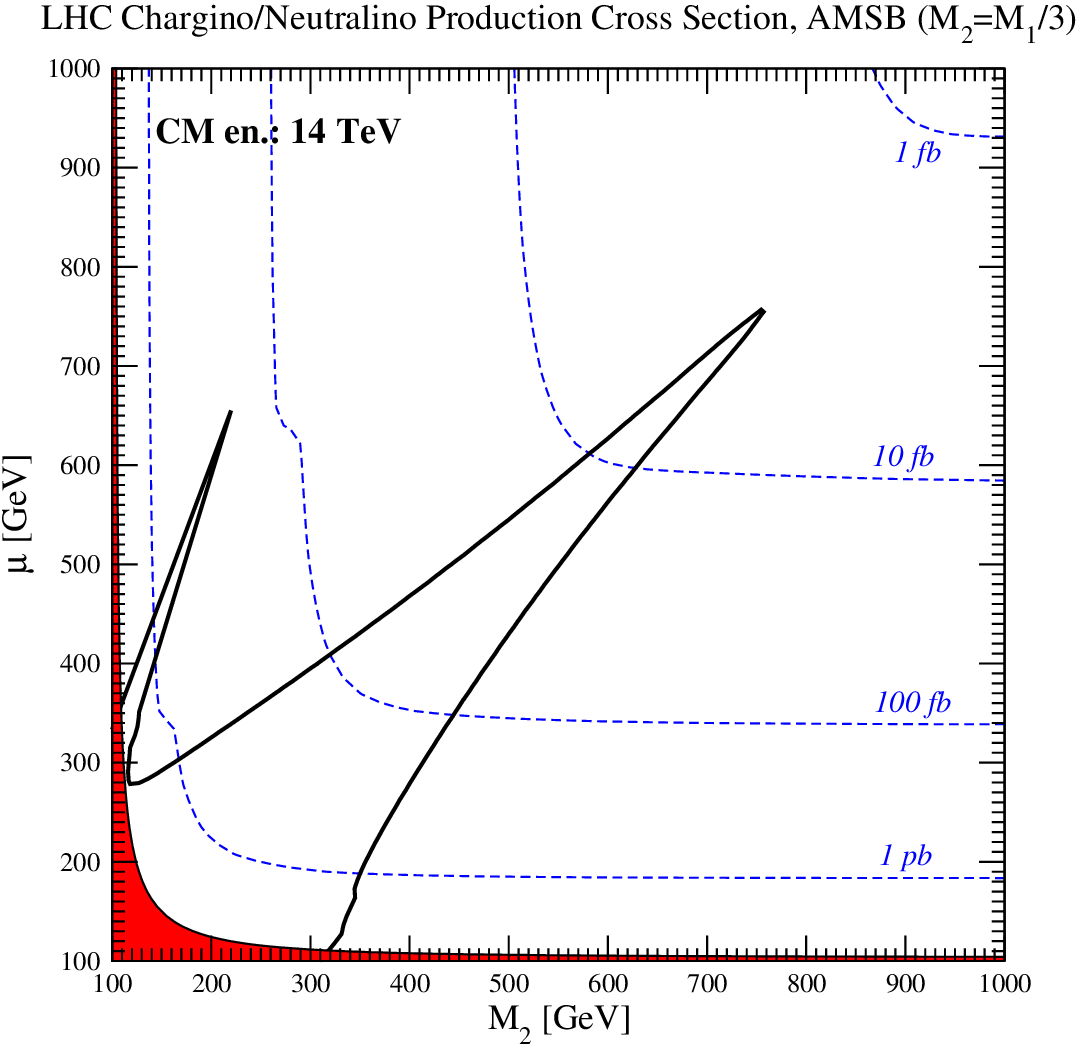}}
\caption{\label{fig:lhc_amsb}\it\small Same as Fig.~\ref{fig:lhc_tb40} in the $(M_2,\mu)$ plane for the AMSB gaugino mass hierarchy}
\end{figure*}

As the neutralinos and charginos in the resonant EWB scenario under consideration are relatively light, with masses typically in the 100 GeV-1 TeV range, one might hope to observe their production from collisions at the Large Hadron Collider (LHC).  All other sparticles in this picture possess masses in the multi-TeV range and are thus not expected to be produced in significant numbers at the LHC. The possible exception is the gluino, which would be naturally expected to have a mass scale comparable to that of the other two gauginos (the wino and bino). The gluino mass is, however, entirely unrelated to the phenomenology of EWB, and is thus essentially a free parameter, making it difficult to ascertain sensible predictions for the LHC based upon gluino production. We therefore only consider the electroweak -inos, i.e. charginos and neutralinos, in what follows.

We calculate here the leading order (LO) total cross-section for electroweak production of neutralinos/charginos at the LHC, with all possible pairs of -inos in the final state, for points in the gaugino-higgsino mass planes suitable for resonant EWB using a modified version of PROSPINO  \cite{Beenakker:1999xh}.  We do so for center-of-mass energies $\sqrt{s}=7$ TeV and 14 TeV, and outline curves of constant total cross-section in the $(M_1,\mu)$ and $(M_2,\mu)$ planes for the gravity-mediated SUSY breaking and AMSB scenarios, respectively.  Our results are displayed in Figs.~\ref{fig:lhc_tb40} and \ref{fig:lhc_amsb}.  For the particle spectrum, we use the values for the masses and MSSM parameters discussed in Sec.~\ref{sec:asymmetry} while calculating the neutralino/chargino masses and their mixing matrices for each point in the $(M_1(M_2),\mu)$ planes.

The various -ino production cross sections approximately follow the particle spectrum, as expected e.g. from the results displayed in Figs.~15-16 of Ref.~\cite{Baer:2005ky},  which pertain to the particular slices of the $(M_1,\mu)$ parameter space with a thermal relic density matching the dark matter density. As in Ref.~\cite{Baer:2005ky} we also find that the largest production cross sections correspond, typically, to the next-to-lightest neutralino plus chargino ($\widetilde{\chi}^0_2\widetilde{\chi}_{1,2}^{\pm}$), and to lightest chargino pairs ($\widetilde{\chi}_1^+\widetilde{\chi}_1^-$); also significant are cross-sections corresponding to $\widetilde{\chi}^0_2 \widetilde{\chi}^0_2$ and $\widetilde{\chi}_2^+\widetilde{\chi}_2^-$ final states. Ref.~\cite{Baer:2005ky} considered in detail the so-called clean trilepton signature \cite{Baer:1994nr}, originating from both chargino-neutralino and from neutralino-neutralino production. After imposing cuts on the lepton transverse momentum, on invariant dilepton masses and a transverse mass veto, and after considering in detail Standard Model backgrounds, Ref.~\cite{Baer:2005ky} concluded that the 5$\sigma$ discovery potential for the LHC with a center of mass energy of 14 TeV and with 100 fb${}^{-1}$ integrated luminosity corresponds to a total -ino production cross  section (before cuts) of about 600 fb.

While other strategies besides the clean trilepton signature with the cuts implemented in Ref.~\cite{Baer:2005ky} are possible and should be pursued, our predicted total -ino production cross sections indicate that the LHC reach in this channel will not exhaust the parameter space covered by dark matter searches. Specifically, we find that a cross section on the order of 600 fb for 14 TeV center of mass collisions will only probe the low-mass end of the EWB-compatible $(M_1,\mu)$ parameter space, namely values of $\mu\lesssim200$ GeV (see Fig.~\ref{fig:lhc_tb40}). Slightly larger values of $\mu$ might be accessible if $M_2<M_1$, but far from exhausting the EWB-compatible parameter space in the wino-resonant funnel, see Fig.~\ref{fig:lhc_amsb}. 

Finally, we note that the study in Ref.~\cite{Baer:2005ky} suggests that the 14 TeV LHC with 10 fb$^{-1}$ of integrated luminosity accesses cross sections $\gtrsim 6$ pb via the clean trilepton channel, with the 7 TeV LHC probing only larger cross sections.  Thus we see from Figs. 6-7 that the current 7 TeV LHC run at the time of writing has so far probed very little of the relevant parameter space for $M_2=2 M_1$ (the corresponding region has in fact already been ruled out by LEP \cite{Amsler:2008zzb}; see e.g. Fig. 1) and only a small sliver of the parameter space near the bottom and left edges of Fig. 7 for the $M_1=3 M_2$ scenario.

\section{Discussion and Conclusions}\label{sec:disc}

In this study we have shown that recent results from both direct and indirect dark matter detection experiments significantly constrain the regions of the MSSM otherwise viable for electroweak baryogenesis.  We have illustrated this for two particular classes of supersymmetry-breaking models, however since our results follow primarily from the relative hierarchy of the neutralinos and charginos and not the precise values  of the individual gaugino masses, we consider our conclusions to be quite general for a given gaugino mass hierarchy satisfying either $M_1<M_2$ or $M_1>M_2$.  In particular, we have found that: 
\begin{itemize} 

\item EWB scenarios with a wino- (or higgsino-, higgsino-wino-) like LSP, occurring when $M_2<M_1$, are essentially ruled out by recent dark matter search null results, unless $R$-parity is violated or the LSP is not the primary DM constituent.  This is due to the sensitivity of direct detection experiments and neutrino telescopes to light neutralinos with significant wino-higgsino mixing combined with the sensitivity of dwarf spheroidal galaxy gamma ray observations to the complementary wino- and higgsino-like LSP parameter space.

\item Although EWB models with a bino-type LSP ($M_1<M_2$) are not entirely ruled out, only a small portion of the neutralino-driven EWB parameter space is still viable for small values of $m_A$ and large $\sin\phi_{\mu}$; for large $m_A$ and/or small CP-violating phase, chargino-driven EWB with a large wino mass becomes the only possibility.  Additionally, with moderate improvements in sensitivity, direct and indirect DM detection results will probe larger portions, and in some cases all, of the $M_1\sim\mu$ funnel and the lower portion of the $M_2\sim \mu$ funnel.  Further, if the LSP is the primary DM candidate, enforcing the correct relic density rules out the whole $M_1\sim \mu$ funnel for maximal CP-violating phase.

\item The LHC will probe only the low-mass portions of the EWB-compatible parameter space for both the $M_2>M_1$ and $M_2<M_1$ scenarios with 100 fb${}^{-1}$ through events with neutralino and chargino pairs in the final state via the clean trilepton signature.  Thus, we expect DM searches to typically provide more stringent constraints on the EWB-compatible regions of the MSSM than the LHC through this channel.
 
\end{itemize}
We emphasize that the conclusions above follow simply from requiring EWB to account for the observed baryon asymmetry and the LSP to be stable and to compose the dark matter density; relaxing these two requirements will weaken our conclusions by potentially extending EWB-compatible regions beyond those considered here and by mitigating the impact of the various dark matter searches on the gaugino-higgsino mass planes.  

It is also worth mentioning that our results may not necessarily apply to all of the so-called ``baryogenesis window" in the MSSM discussed in Ref.~\cite{Carena:2008vj}, as the authors considered the case of $m_A$ in the multi-TeV range and thus relied on sources not considered in the present study, being subdominant for small to moderate $m_A$ and large $\tan \beta$.  However, portions of the gaugino-higgsino mass planes with a heavy pseudoscalar $A$ boson may still be probed with e.g. neutrino telescopes, whose sensitivity increases for large $m_A$ (see Fig.~\ref{fig:tb40_2}).  Also, the choice of a heavy CP-odd Higgs is not fundamental to the light stop scenario and when this condition is relaxed, we expect our conclusions to constrain much of the allowed parameter space.  In particular, requiring the RH stop to be light implies only a small portion of the gaugino-higgsino mass planes in the lower left hand corner consistent with a neutralino LSP, and this corner of the parameter space is precisely the one most tightly constrained by DM results and expected to be probed by the LHC.  We leave a detailed study of the applicability of our results to the case of a heavy $A$ and non-resonant sources to future work.

The present study also implicitly guides the answer to the question: what could one conclude regarding electroweak baryogenesis should a {\em positive} dark matter detection signal arise? For example, should the next generation of direct detection experiments detect uncontroversial signals of WIMP scattering, we could argue, from Fig.~\ref{fig:tb40_1} that successful electroweak baryogenesis would force one into either the tip of the $M_1\sim\mu$ funnel, or into the bulk of the $M_2\sim\mu$ region. In turn, Fig.~\ref{fig:tb40_2} would then imply a signal at IceCube plus DeepCore in the latter case, and no signal in the former and, depending upon the value of $m_A$, possibly a signal from Fermi. Vice-versa, a signal from IceCube/DeepCore could signal that, if this setup is the correct description of how the baryon asymmetry is generated, one should get a positive detection with ton-sized direct detection experiments

In our determination of the parameter space producing the correct observed baryon asymmetry we have been careful to be conservative, addressing the dependence on the various model uncertainties and choosing optimistic values to maximize the EWB compatible regions.  As a result, although many details of the MSSM parameter space have yet to be tightly constrained, we expect our conclusions to remain applicable as continued progress is made in this regard.

In conclusion, the window for electroweak baryogenesis in the MSSM is closing fast.  Barring the discovery of either the particle constituting the dark matter and/or supersymmetry in the near future, one may be left with only a small portion of the entire MSSM parameter space viable simultaneously for EWB and a neutralino LSP dark matter candidate, leading one to rely on different sources or extended models to explain the matter content of the universe we observe today.

\begin{acknowledgments}
\noindent  SP is partly supported by an Outstanding Junior Investigator Award from the US Department of Energy and by Contract DE-FG02-04ER41268, and by NSF Grant PHY-0757911. 
\end{acknowledgments}


\begin{thebibliography}{300}

 \bibitem{Huet:1995sh}
  P.~Huet, A.~E.~Nelson,
  Phys.\ Rev.\  {\bf D53}, 4578-4597 (1996).
  [hep-ph/9506477].

\bibitem{EWB_and_EDMs}
  V.~Cirigliano, Y.~Li, S.~Profumo and M.~J.~Ramsey-Musolf,
  JHEP {\bf 1001}, 002 (2010)
  [arXiv:0910.4589 [hep-ph]].

  \bibitem{EWB_and_DM}
  V.~Cirigliano, S.~Profumo and M.~J.~Ramsey-Musolf,
  JHEP {\bf 0607}, 002 (2006)
  [arXiv:hep-ph/0603246].
  
   \bibitem{Resonant_Relaxation}
  C.~Lee, V.~Cirigliano and M.~J.~Ramsey-Musolf,
  Phys.\ Rev.\  D {\bf 71}, 075010 (2005)
  [arXiv:hep-ph/0412354].
  
  \bibitem{Chung:2008aya}
  D.~J.~H.~Chung, B.~Garbrecht, M.~J.~Ramsey-Musolf, S.~Tulin,
  Phys.\ Rev.\ Lett.\  {\bf 102}, 061301 (2009).
  [arXiv:0808.1144 [hep-ph]].
  
   \bibitem{Lepton_Mediated}
  D.~J.~H.~Chung, B.~Garbrecht, M.~J.~Ramsey-Musolf and S.~Tulin,
  Phys.\ Rev.\  D {\bf 81}, 063506 (2010)
  [arXiv:0905.4509 [hep-ph]].
  
  \bibitem{Supergauge}
  D.~J.~H.~Chung, B.~Garbrecht, M.~J.~Ramsey-Musolf and S.~Tulin,
  Phys.\ Rev.\ Lett.\  {\bf 102}, 061301 (2009)
  [arXiv:0808.1144 [hep-ph]].

  \bibitem{Including_Yukawa}
  V.~Cirigliano, M.~J.~Ramsey-Musolf, S.~Tulin and C.~Lee,
  Phys.\ Rev.\  D {\bf 73}, 115009 (2006)
  [arXiv:hep-ph/0603058].

  \bibitem{Carena:2002ss}
  M.~S.~Carena, M.~Quiros, M.~Seco, C.~E.~M.~Wagner,
  Nucl.\ Phys.\  {\bf B650}, 24-42 (2003).
  [hep-ph/0208043].
  
   \bibitem{Carena:2008vj}
  M.~Carena, G.~Nardini, M.~Quiros, C.~E.~M.~Wagner,
  Nucl.\ Phys.\  {\bf B812}, 243-263 (2009).
  [arXiv:0809.3760 [hep-ph]].
  
  \bibitem{Konstandin:2004gy}
  T.~Konstandin, T.~Prokopec, M.~G.~Schmidt,
  Nucl.\ Phys.\  {\bf B716}, 373-400 (2005).
  [hep-ph/0410135].
  
  \bibitem{Konstandin:2003dx}
  T.~Konstandin, T.~Prokopec and M.~G.~Schmidt,
  Nucl.\ Phys.\  B {\bf 679}, 246 (2004)
  [arXiv:hep-ph/0309291].

   \bibitem{More_Relaxed}
  A.~Riotto,
  Phys.\ Rev.\  D {\bf 58}, 095009 (1998)
  [arXiv:hep-ph/9803357].
  
    \bibitem{Balazs:2004ae}
  C.~Balazs, M.~S.~Carena, A.~Menon, D.~E.~Morrissey, C.~E.~M.~Wagner,
  Phys.\ Rev.\  {\bf D71}, 075002 (2005).
  [hep-ph/0412264].
  
    \bibitem{Huber:2001xf}
  S.~J.~Huber, P.~John, M.~G.~Schmidt,
  Eur.\ Phys.\ J.\  {\bf C20}, 695-711 (2001).
  [hep-ph/0101249].
  
  \bibitem{CPV_and_EWB}
  M.~S.~Carena, J.~M.~Moreno, M.~Quiros, M.~Seco and C.~E.~M.~Wagner,
  Nucl.\ Phys.\  B {\bf 599}, 158 (2001)
  [arXiv:hep-ph/0011055].
  
  \bibitem{Cline:2000kb}
  J.~M.~Cline and K.~Kainulainen,
  Phys.\ Rev.\ Lett.\  {\bf 85}, 5519 (2000)
  [arXiv:hep-ph/0002272].
     
  \bibitem{Komatsu:2010fb}
  E.~Komatsu {\it et al.}  [WMAP Collaboration],
  Astrophys.\ J.\ Suppl.\  {\bf 192}, 18 (2011)
  [arXiv:1001.4538 [astro-ph.CO]].
  
  \bibitem{Baer:2000gf}
  H.~Baer, M.~A.~Diaz, P.~Quintana and X.~Tata,
  JHEP {\bf 0004}, 016 (2000)
  [arXiv:hep-ph/0002245].
   
   \bibitem{Amsler:2008zzb}
  C.~Amsler {\it et al.}  [Particle Data Group],
  Phys.\ Lett.\  B {\bf 667}, 1 (2008).
  
  \bibitem{Laine:1998qk}
  M.~Laine and K.~Rummukainen,
  Nucl.\ Phys.\  B {\bf 535}, 423 (1998)
  [arXiv:hep-lat/9804019].

  \bibitem{Laine:2000xu}
  M.~Laine,
  arXiv:hep-ph/0010275.
  
  \bibitem{Carena:2008rt}
  M.~Carena, G.~Nardini, M.~Quiros, C.~E.~M.~Wagner,
  JHEP {\bf 0810}, 062 (2008).
  [arXiv:0806.4297 [hep-ph]].
  
\bibitem{Heavy_Stops} See e.g. M.~Pietroni,
  Nucl.\ Phys.\  {\bf B402}, 27-45 (1993).
  [hep-ph/9207227] and references therein; 
  R.~Fok and G.~D.~Kribs,
  Phys.\ Rev.\  D {\bf 78}, 075023 (2008)
  [arXiv:0803.4207 [hep-ph]].
  and references therein; 
  J.~Shu, T.~M.~P.~Tait and C.~E.~M.~Wagner,
  Phys.\ Rev.\  D {\bf 75}, 063510 (2007)
  [arXiv:hep-ph/0610375].
  and references therein; 
  P.~Fileviez Perez, T.~Han, G.~y.~Huang, T.~Li and K.~Wang,
  Phys.\ Rev.\  D {\bf 78}, 015018 (2008)
  [arXiv:0805.3536 [hep-ph]].
  
  \bibitem{Profumo:2007wc}
  S.~Profumo, M.~J.~Ramsey-Musolf, G.~Shaughnessy,
  JHEP {\bf 0708}, 010 (2007).
  [arXiv:0705.2425 [hep-ph]].

  \bibitem{DeSimone:2011ek}
  A.~De Simone, G.~Nardini, M.~Quiros and A.~Riotto,
  arXiv:1107.4317 [hep-ph].
  
   \bibitem{Giudice:1993bb}
  G.~F.~Giudice, M.~E.~Shaposhnikov,
  Phys.\ Lett.\  {\bf B326}, 118-124 (1994).
  [hep-ph/9311367].
  
 \bibitem{Precision_EW1}
  E.~Carmona  [ANTARES Collaboration],
  Nucl.\ Phys.\ Proc.\ Suppl.\  {\bf 95}, 161 (2001).
  M.~S.~Carena, M.~Quiros and C.~E.~M.~Wagner,
  Nucl.\ Phys.\  B {\bf 524}, 3 (1998)
  [arXiv:hep-ph/9710401].
  J.~M.~Cline, M.~Joyce and K.~Kainulainen,
  Phys.\ Lett.\  B {\bf 417}, 79 (1998)
  [Erratum-ibid.\  B {\bf 448}, 321 (1999)]
  [arXiv:hep-ph/9708393].
    
   \bibitem{vw}
  G.~D.~Moore,
  JHEP {\bf 0003}, 006 (2000)
  [arXiv:hep-ph/0001274].
  P.~John and M.~G.~Schmidt,
  Nucl.\ Phys.\  B {\bf 598}, 291 (2001)
  [Erratum-ibid.\  B {\bf 648}, 449 (2003)]
  [arXiv:hep-ph/0002050].
  P.~John and M.~G.~Schmidt,
  arXiv:hep-ph/0012077.
  A.~Megevand and A.~D.~Sanchez,
  Nucl.\ Phys.\  B {\bf 825}, 151 (2010)
  [arXiv:0908.3663 [hep-ph]].

  \bibitem{MSSM_Bubbles}
  J.~M.~Moreno, M.~Quiros and M.~Seco,
  Nucl.\ Phys.\  B {\bf 526}, 489 (1998)
  [arXiv:hep-ph/9801272].

  \bibitem{Li:2008ez}
  Y.~Li, S.~Profumo, M.~Ramsey-Musolf,
  Phys.\ Lett.\  {\bf B673}, 95-100 (2009).
  [arXiv:0811.1987 [hep-ph]].

 \bibitem{Bornhauser:2010mw}
  S.~Bornhauser, M.~Drees, S.~Grab, J.~S.~Kim,
  Phys.\ Rev.\  {\bf D83}, 035008 (2011).
  [arXiv:1011.5508 [hep-ph]].
  
  \bibitem{Aprile:2011hi}
  E.~Aprile {\it et al.} [ XENON100 Collaboration ],
  [arXiv:1104.2549 [astro-ph.CO]].
  
  \bibitem{elenaquote}
E.~Aprile, private communication.

  \bibitem{icecube}
 A.~M.~Brown and o.~b.~o.~Collaboration,
  arXiv:1012.1633 [astro-ph.HE];
 R.~Abbasi {\it et al.} [ ICECUBE Collaboration ],
  Phys.\ Rev.\ Lett.\  {\bf 102}, 201302 (2009).
  [arXiv:0902.2460 [astro-ph.CO]].
  
  \bibitem{dsph}
  A.~A.~Abdo, M.~Ackermann, M.~Ajello, W.~B.~Atwood, L.~Baldini, J.~Ballet, G.~Barbiellini, D.~Bastieri {\it et al.},
  Astrophys.\ J.\  {\bf 712}, 147-158 (2010).
  [arXiv:1001.4531 [astro-ph.CO]].

\bibitem{amsb}
A.~Pomarol, R.~Rattazzi,
  JHEP {\bf 9905}, 013 (1999).
  [hep-ph/9903448];
  J.~L.~Feng, T.~Moroi,
  Phys.\ Rev.\  {\bf D61}, 095004 (2000).
  [hep-ph/9907319];
T.~Moroi, L.~Randall,
  Nucl.\ Phys.\  {\bf B570}, 455-472 (2000).
  [hep-ph/9906527].
    
  \bibitem{quintess}
   S.~Profumo, P.~Ullio,
  JCAP {\bf 0311}, 006 (2003).
  [hep-ph/0309220].
  
  \bibitem{Beenakker:1999xh}
  W.~Beenakker, M.~Klasen, M.~Kramer, T.~Plehn, M.~Spira and P.~M.~Zerwas,
  Phys.\ Rev.\ Lett.\  {\bf 83}, 3780 (1999)
  [Erratum-ibid.\  {\bf 100}, 029901 (2008)]
  [arXiv:hep-ph/9906298].

  \bibitem{Baer:2005ky}
  H.~Baer, T.~Krupovnickas, S.~Profumo, P.~Ullio,
  JHEP {\bf 0510}, 020 (2005).
  [hep-ph/0507282].

  \bibitem{Baer:1994nr}
  H.~Baer, C.~h.~Chen, F.~Paige and X.~Tata,
  Phys.\ Rev.\  D {\bf 50} (1994) 4508
  [arXiv:hep-ph/9404212].
  
  


\end{thebibliography}
\end{document}